\documentclass[a4paper,fleqn,usenatbib]{mnras}

\usepackage{txfonts}

\usepackage[T1]{fontenc}
\usepackage{ae,aecompl}

\usepackage{graphicx}	
\usepackage{amstext}
\usepackage{amssymb}	

\newcommand{\kms}{$\rm km~s^{-1}$}

\newcommand{\zgal}{CGCG097-}

\title[MUSE goes BIG]{MUSE sneaks a peek at extreme ram-pressure stripping events - IV.  Hydrodynamic and gravitational interactions in the Blue Infalling Group}

\author[Fossati et al.]{Matteo Fossati$^{1,2,3}$\thanks{E-mail: matteo.fossati@durham.ac.uk}, 
Michele Fumagalli$^{1,2}$, 
Giuseppe Gavazzi$^{4}$,
Guido Consolandi$^{4}$,
\newauthor
Alessandro Boselli$^{5}$,
Masafumi Yagi${^6}$,
Ming Sun${^7}$,
David J. Wilman$^{8,3}$
\\
  $^{1}$Centre for Extragalactic Astronomy, Durham University, South Road, Durham, DH1 3LE, UK \\
  $^{2}$Institute for Computational Cosmology, Durham University, South Road, Durham, DH1 3LE, UK \\
  $^{3}$Max-Planck-Institut f\"{u}r Extraterrestrische Physik, Giessenbachstrasse, 85748, Garching, Germany \\
  $^{4}$Universit\'a di Milano-Bicocca, Piazza della Scienza 3, 20100, Milano, Italy \\
  $^{5}$Aix Marseille Universit{\'e}, CNRS, LAM, Laboratoire d'Astrophysique de Marseille, Marseille, France \\
  $^{6}$Optical and Infrared Astronomy Division, National Astronomical Observatory of Japan, Mitaka, Tokyo, 181-8588, Japan \\
  $^{7}$Department of Physics and Astronomy, University of Alabama in Huntsville, Huntsville, AL 35899, USA \\
  $^{8}$Universit{\"a}ts-Sternwarte M{\"u}nchen, Scheinerstrasse 1,81679 M{\"u}nchen, Germany\\
 }

\date{Accepted 2019 January 9. Received 2019 January 9; in original form 2018 December 20.}

\pubyear{\the\year}

\begin{document}
\label{firstpage}
\pagerange{\pageref{firstpage}--\pageref{lastpage}}
\maketitle

\begin{abstract}
We report new wide-field ($\approx 4\times 4$ arcmin$^2$) MUSE observations 
of the Blue Infalling Group (BIG), a compact group of galaxies 
located at a projected distance of $\simeq 150$ kpc from the X-Ray centre of the A1367 cluster at $z=0.021$.
Our MUSE observations map in detail the extended ionized gas, primarily traced by H$\alpha$ emission, in between the members of the group. The gas morphology and its kinematics appear consistent with 
a tidal origin due to galaxy encounters, as also supported by the disturbed kinematics visible in one of the group members and the presence of tidal dwarf systems. A diffuse tail extending in the direction opposite to the cluster centre is also detected, hinting at a global ram-pressure stripping of the intra-group material as BIG falls inside A1367. Based on the analysis of spatially-resolved emission line maps, we identify multiple ionization mechanisms for the diffuse gas filaments, including {\it in situ} photoionization from embedded HII regions and shocks. Combining spatially resolved kinematics and line ratios, we rule out the association of the most massive galaxy, \zgal120, with the group as this system appears to be decoupled from the intragroup medium and subject to strong ram pressure as it falls into A1367. Through our new analysis, we conclude that BIG is shaped by pre-processing produced by gravitational interactions in the local group environment combined with ram pressure stripping by the global cluster halo.
\end{abstract}

\begin{keywords}
galaxies: clusters: intracluster medium -- galaxies: evolution -- galaxies: interactions -- galaxies: groups: individual: A1367 -- ISM: evolution
\end{keywords}



\section{Introduction}
Galaxies are distributed within the universe in a non-uniform way, spanning a wide range of densities from the core of rich clusters, to compact and loose groups, filaments and voids. Starting from the seminal work of \citet{Dressler80}, it became evident that the environment plays a major role in shaping galaxy evolution. Large scale multiwavelength surveys \citep[e.g. the Sloan Digital Sky Survey, ][]{York00} have provided a huge momentum for studies of the environment. We have now a clear picture that massive haloes (i.e. clusters of galaxies) are mainly composed of quiescent early-type objects, while the lower density field is populated by star-forming late-type systems, and that this picture emerges already when the Universe was less than half of its current age \citep{Dressler97, Kauffmann03, Baldry06, Gavazzi10, Peng10, Wetzel13, Fossati17}. Several physical processes have been proposed in the literature to explain the observed differences between cluster and field galaxies, as reviewed in \citet{Boselli06}, including gravitational interactions with other cluster or group members or with the halo potential \citep{Merritt83, Moore98, Byrd90}, and the hydrodynamic interactions between the galaxy interstellar medium (ISM) and the intra-cluster medium (ICM), including ram pressure stripping \citep{Gunn72}. 

Several modern multi-wavelength surveys in nearby clusters identify ram-pressure stripping as the leading mechanism for the quenching of dwarf galaxies, which are still infalling at high rate onto massive clusters today \citep{Gavazzi10, Boselli16a, Oman16}. Conversely, early-type massive galaxies in clusters are thought to have formed at early epochs, by violent starbursts, possibly driven by merging events or gravitational interactions \citep{Thomas05}. Recent emphasis has been given to ``pre-processing'' \citep{Fujita04, Dressler04}, that is the cumulative effect of environmental transformations that occur within intermediate-mass groups, before they are accreted onto larger haloes like galaxy clusters. In contrast to ram pressure, pre-processing is emerging as the key player in the early stages of assembly of dynamically-young clusters, especially at moderate redshifts \citep{Dressler13, Mok13}. \citet{Wetzel13}, studying a large statistical sample of galaxies in the local Universe found that $\simeq 30\%$ of the galaxies in present-day clusters were quenched in another halo (e.g. in lower mass groups), and that this fraction mildly depends on stellar mass. However, the physical processes responsible for the evolution of galaxies from star forming to passive systems in different environments and at different epochs are under debate, and this constitutes to be a major quest of modern studies on galaxy evolution.

In order to study the physics of pre-processing we need to observe groups of galaxies on their way towards accretion on a massive cluster. The Blue Infalling Group \citep[BIG][]{Sakai02, Cortese06} is an ideal candidate due to its proximity. Named BIG by \citet{Gavazzi03a}, the group features 3 galaxies more massive than $10^9\rm{M_\odot}$, and a number of lower-mass star-forming systems connected by a complex of ionized gas filaments discovered in deep H$\alpha$ emission line images by \citet{Cortese06}. The group is located $\approx 150$ kpc from one of the subclusters that forms the A1367 cluster \citep{Cortese04}, and has a mean recessional velocity of $v \simeq 8230$ \kms\ \citep{Gavazzi03a}. The cluster has instead a mean velocity of $v\simeq6484$ \kms\ \citep{Cortese04}, which suggested that BIG is currently falling into the cluster potential. This scenario is supported by the presence of a $\simeq 330$ kpc long ionized gas tail extending in the North-West direction, possibly opposite to the direction of motion of the group in the cluster environment \citep{Yagi17}.

As seen in projection, the group is composed of three massive galaxies \zgal125 ($\log(M_*/{\rm M_\odot})=10.44$), \zgal114 ($\log(M_*/{\rm M_\odot})=9.29$), and \zgal120 ($\log(M_*/{\rm M_\odot})=10.84$), located at the vertices of a triangle with a projected side of $\simeq 45$ kpc\footnote{The stellar masses are obtained using the \citet{Zibetti09} mass to light ratios using SDSS $g-$ and $i-$band magnitudes.}. While \zgal125 and \zgal114 have a similar recessional velocity, consistent with that of the group, \zgal120 has a much lower velocity ($v\simeq5620$ \kms), which questions its membership within the group. However the tantalising presence of filaments of gas surrounding \zgal120 opened a debate about the role of this galaxy within BIG. \citet{Cortese06} presented a significant amount of observational material, including deep H$\alpha$ images that revealed the ionized gas filaments, optical long slit spectroscopy of the galaxies and the star forming dwarf systems and atomic hydrogen observations with the Arecibo radio telescope. Their data revealed that \zgal125 is the result of a galaxy merger that could have produced the ionized gas filaments. However, the role of \zgal120 remained unclear. To answer this question and to understand in details the timescales and the role of different physical mechanisms in the group environment we need deep however resolved spectroscopic observations to unveil the kinematics and physical conditions of the ionized gas, and the galaxy stellar kinematics in their full extent.

In this work we present a large mosaic ($\approx 4\times 4$ arcmin$^2$) of optical integral field observations of BIG taken in 13 pointings of the Multi Unit Spectroscopic Explorer \citep[MUSE;][]{Bacon10} mounted on UT4 of the ESO Very large Telescope. 
The superior sensitivity, image quality, field of view (1$\times$1 arcmin) and spectral bandwidth ($\sim4800-9300\AA$) of MUSE, makes it the best instrument to produce mosaics of extended objects like a group environment in the local Universe.

In Section \ref{obs_datared} we present an improved data reduction scheme to handle the data volume of the large mosaic, while in Section \ref{sec_fitting} we describe the techniques used to derive the stellar continuum and emission line maps. We then present the study of the gas and stellar kinematics in Section \ref{gas_star_kin}, and of the gas ionization conditions in Section \ref{emi_diagnostics}. We discuss the implications of new and previous observations for the evolutionary history of BIG in Section \ref{discussion}, while our conclusions can be found in Section \ref{sec_conclusions}. 

Throughout, we assume a flat $\Lambda$CDM Universe with $\Omega_M$ = 0.3, $\Omega_\Lambda$ = 0.7, and $H_0$ = 70 $\rm km~s^{-1}~Mpc^{-1}$. With the adopted cosmology, the comoving radial distance of A1367 becomes 92.2 Mpc, 1 arcsec on the sky corresponds to 0.44 kpc, and one arcmin to 26.4 kpc. Where necessary, we adopt a \citet{Chabrier03} stellar initial mass function.

\section{Observations and data reduction} \label{obs_datared}

\begin{figure*}
\centering
\includegraphics[width=0.95\textwidth]{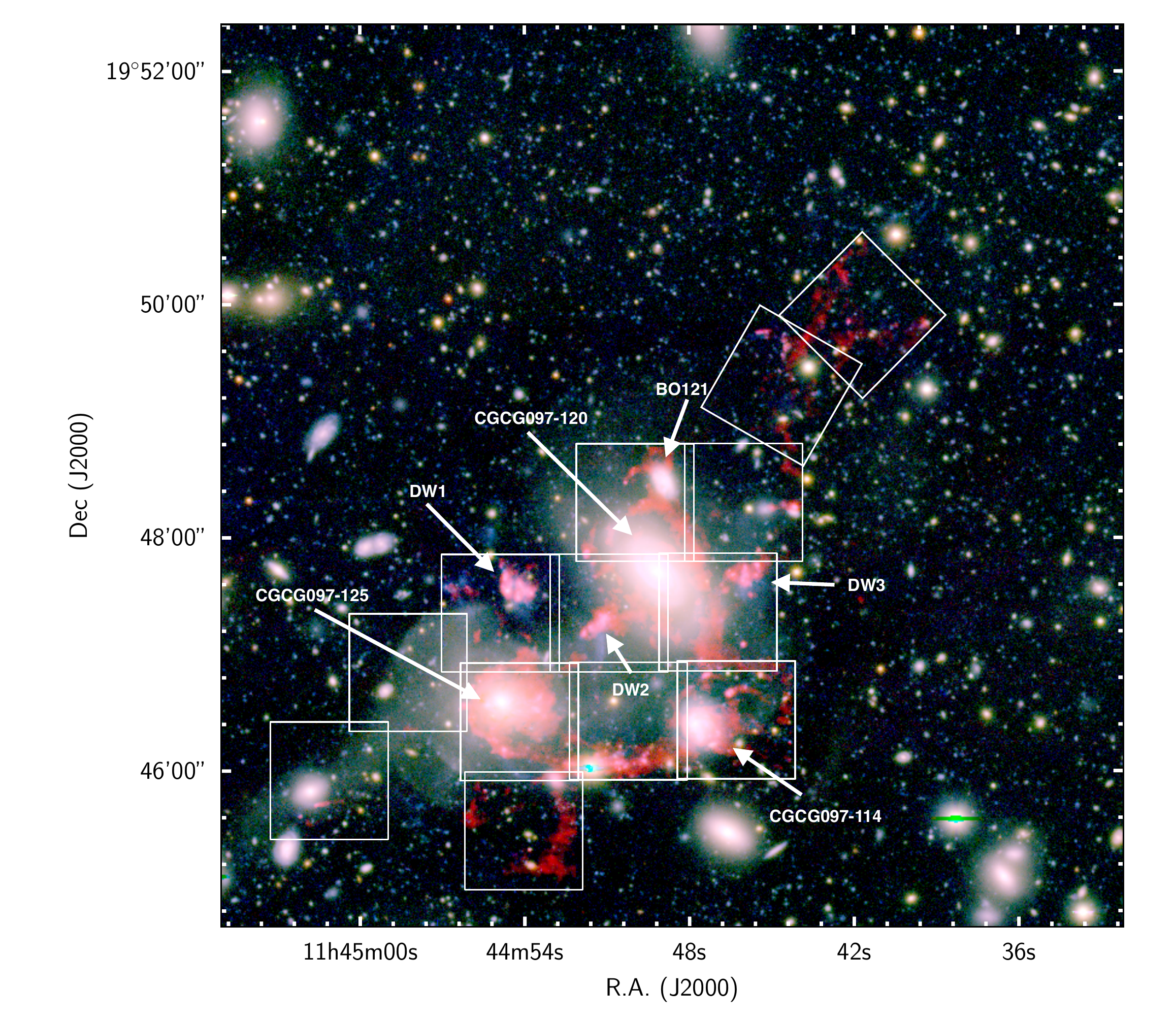}
\caption{RGB ($i,R,B$) image of the region of Abell 1367 containing the BIG group based on broad-band imaging observations by \citet{Yagi17}. The H$\alpha$ flux map from MUSE data is overlayed in shades of red. The white lines indicate the position of the 13 individual MUSE pointings. The names of objects in the field, as defined in \citet{Cortese06}, are shown in white.}
\label{BIG_RGB_ima}
\end{figure*}

MUSE observations have been completed as part of the ESO programme ID 095.B-0023 (PI Fumagalli) between April 12 2015 and February 9 2016, using the wide field mode and the nominal wavelength range instrument setup. We observed BIG in mosaic, covering a total area of $\simeq 13~\rm arcmin^2$. The core of the group was observed in 10 pointings distributed in a squared region of $\simeq 3~\rm arcmin$ on a side. In addition, we targetted the extended ionized gas tail with two MUSE pointings and an additional pointing has been placed on a dwarf cluster galaxy on the East side of the group. Observations have been taken under clear skies, with good image quality in the range FWHM $\simeq 0.7-1.1$ arcsec. Figure \ref{BIG_RGB_ima} shows an color image of the region of A1367 containing the BIG group. The individual footprints of MUSE pointings are shown in white.

Each pointing was observed with a sequence of $2\times 1200~\rm s$ exposures on target, plus a shorter $240~\rm s$ exposure in a nearby empty sky region. In between exposures, we rotated the instrument by 90 degree and performed small dithers to minimize signatures arising from the different MUSE spectrographs in the final data. 

Data have been reduced with the standard MUSE pipeline \citep[v2.0.1,][]{Weilbacher14}, which performs basic calibrations including bias subtraction, flat-fielding and wavelength plus flux calibrations. At the end of the standard reduction, we reconstruct each exposure into a datacube and we define an absolute astrometric solution matching point-like sources with a catalogue drawn from the SDSS-DR11 database \citep{Alam15}. Then we use the derived World Coordinate System (WCS) offsets to project each individual exposure onto a reference grid that encompasses the entire mosaic with 0.2 arcsec spatial sampling. With this method we ensure that no additional interpolation of the data is required to combine multiple exposures. 

We further process each individual projected cube to reduce residual differences in the illumination of individual integral field units as described in \citet{Consolandi17}. We suppress the residuals of sky lines using the {\sc ZAP} code \citep[v1.0,][]{Soto16}, using single value decomposition tables derived from the external sky exposures. After this step, we collapse the sky subtracted cubes along the wavelength axis and we derive a residual background value using iterative sigma clipping after having masked bright sources in each exposure. We then subtract this value to ensure that the global background in the final mosaic is flat and consistent with zero. Lastly we combine the exposures into the final mosaic using mean statistics and propagating the uncertainties in the variance extension. 

In order to test the flux calibration reached in the final mosaic, we compared the $r$-band flux of few point-like sources with their values in the SDSS-DR11 database, finding consistency within $\simeq 5\%$.

\section{Stellar continuum and emission line fitting} \label{sec_fitting}
Before fitting the ionized gas emission lines, we smooth the datacube with a 10$\times$10 pixels (2$\times$2 arcsec) median filter to improve the signal to noise ratio of individual pixels without compromising the spatial resolution. Hereafter, all the data processing is applied to the smoothed cubes unless otherwise stated.

\begin{figure*}
\centering
\includegraphics[width=0.90\textwidth]{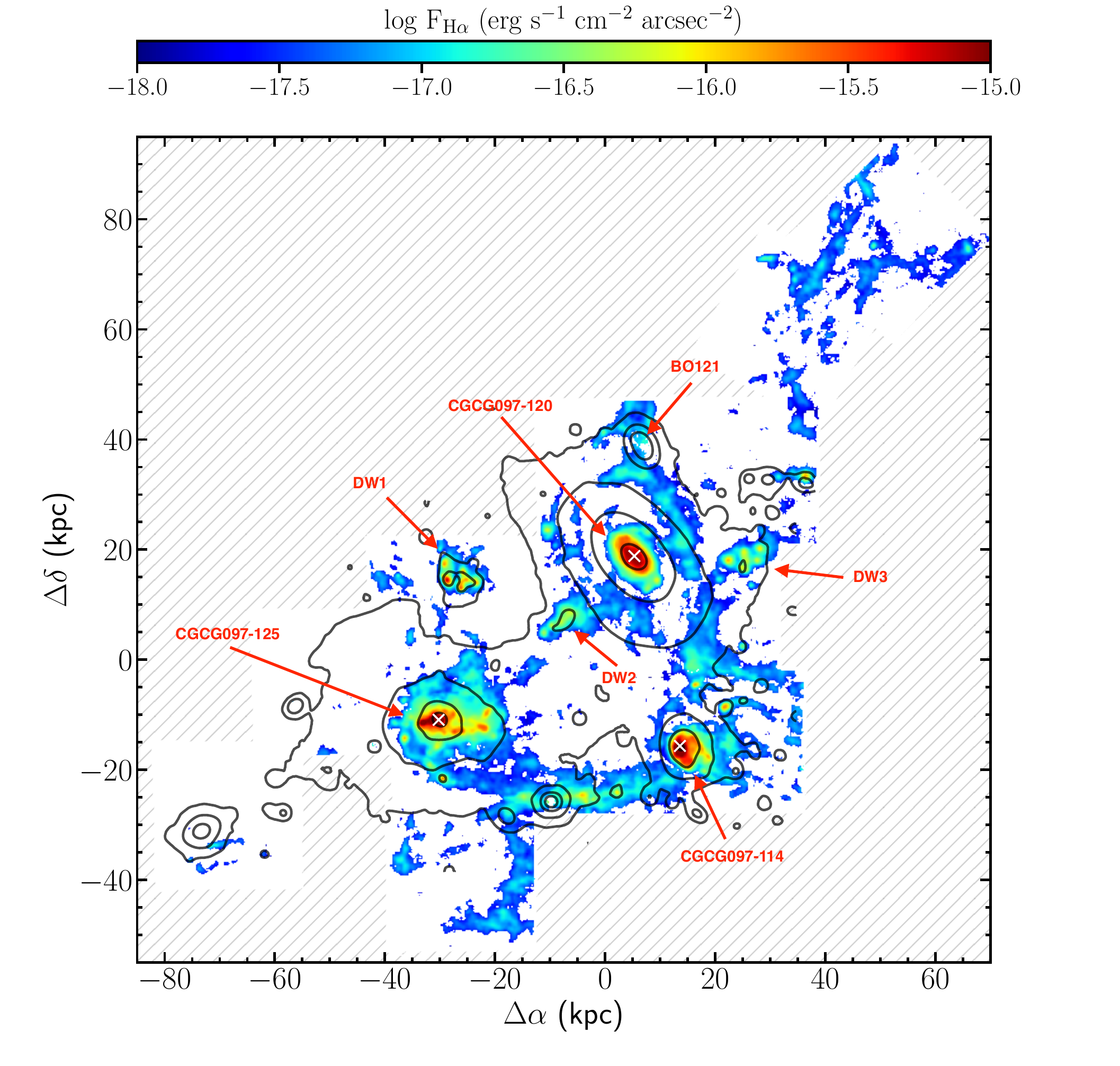}
\caption{H$\alpha$ map of the ionized gas in the BIG group. The grey contours are drawn from the the $R$-band Subaru image at the 20th, 22nd, 24th, and 26th mag/arcsec$^2$ and clearly highlight the large amount of diffuse intra-group light connecting all the main galaxies. Areas not covered by MUSE observations are shaded in grey. The names of objects in the field, as defined in \citet{Cortese06}, are shown in red. The white crosses mark the optical centers of the galaxies.}
\label{Halpha_map}
\end{figure*}

Furthermore, it is necessary to
correct for the stellar absorption underlying the Hydrogen Balmer lines which are present in the MUSE spectra, namely H$\alpha$ and H$\beta$. Similarly to other papers of this series \citep{Fossati16, Consolandi17} we used the {\sc GANDALF} \citep{Sarzi06} code for this purpose. Given the size of the final mosaic, we run {\sc GANDALF} only on cutouts centered on the main galaxies in the field: \zgal125, \zgal114, \zgal120. The galaxy on the NW of \zgal120, named BO121 in \citet{Cortese06}, is a background object with recessional velocity of $v\simeq20300$ \kms, but we also subtract its stellar continuum using {\sc GANDALF} to optimally separate its signal from the ionized gas from BIG.  
{\sc GANDALF} runs in combination with the penalized pixel-fitting code \citep[pPXF; ][]{Cappellari04} to model the stellar continuum  in individual spaxels where the continuum $S/N >$ 15. Due to the extended wavelength range of the MUSE data, we use the stellar spectra without wavelength gaps from the Indo-U.S. Library of Coude Feed Stellar Spectra library \citep{Valdes04}, which optimally covers the MUSE wavelengths at a resolution of $\simeq 1\AA$ FWHM.
In each spaxel, the best fit stellar continuum spectrum is then subtracted from the datacube to obtain a new cube of pure emission lines. As a by-product of this procedure we obtain maps of the stellar kinematics for each galaxy.

In order to fit emission lines, we use the {\sc KUBEVIZ} code \citep[v2.0;][]{Fossati16}. This code fits several groups of emission lines (linesets) using Gaussian templates. Since in the optical spectra, the H$\alpha$ line is usually the strongest, we run the code such that first we fit the H$\alpha$+[NII] lineset, with free kinematics (velocity and dispersion of the lines) and then we fix the kinematics for the fits of the other linesets. 
In order to avoid contamination of the fits due to the background galaxy BO121, we fit this object first and we subtracted the emission of its lines from the datacube.
We then proceded to fit the entire MUSE datacube with {\sc KUBEVIZ}. After the fit, spaxels with $S/N < 5$ in H$\alpha$ or
with uncertainties on the kinematics greater than 50 \kms\ are flagged as bad fits. However, a small fraction of these bad fits
might be caused by a poor choice of the initial guesses. To overcome this issue we take advantage of a new feature of the {\sc KUBEVIZ} code called FITADJ. When this fitting mode is selected, the code fits bad spaxels using the average values from the fits of nearby good spaxels as initial guesses. The code starts
from bad fits which are surrounded by the highest number of good fits and iteratively re-evaluates the new fits against the flagging criteria, gradually moving to more isolated bad fits until no further improvement to the fit map is found. This method increases the number of good fits of the order of $5\%$ on the MUSE map of BIG.

\section{Gas and stellar kinematics} \label{gas_star_kin}
The complete spatial coverage of MUSE allows us to derive maps of the gas distribution and kinematics, therefore extending the long slit spectroscopic measurements of \citet{Cortese06}. In Figure \ref{Halpha_map} we show the H$\alpha$ map in the region covered by MUSE pointings. The grey contours are drawn from the Subaru $R$-band observations by \citet{Yagi17}. We readily notice a spectacular complex of ionized gas which surrounds the three massive galaxies and extends in a long filamentary tail in the N-W direction. These features have been first discovered by \citet{Cortese06} using narrow-band imaging techniques and later re-observed with higher depth and spatial resolution by \citet{Yagi17}. With MUSE, however, we can for the first time derive a full kinematic map of these faint filaments. 

\begin{figure*}
\centering
\includegraphics[width=0.90\textwidth]{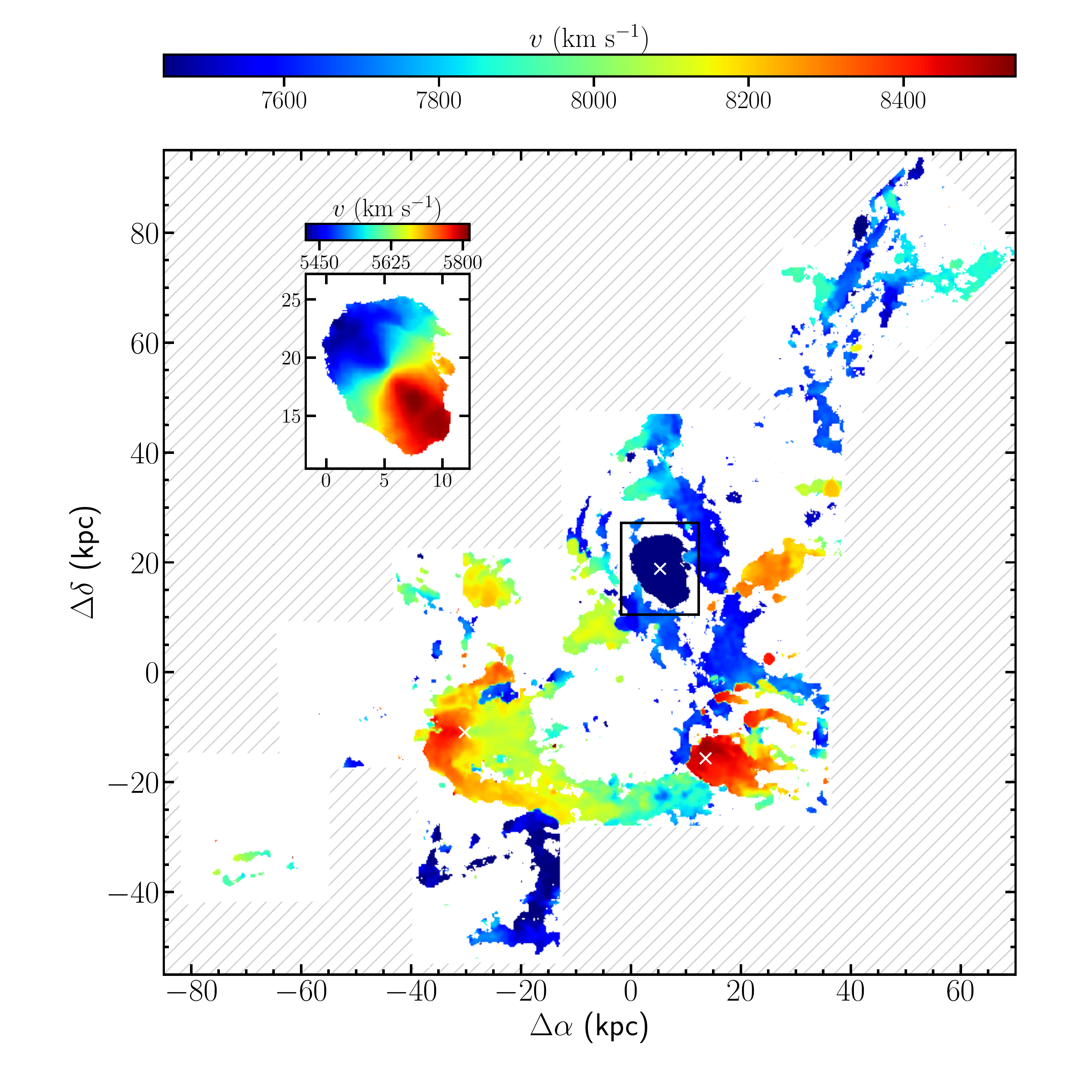}
\caption{Recessional velocity map of the ionized gas from the H$\alpha$+[NII]. Because of the significantly different velocity of the gas in the core of \zgal120, we show a zoomed version of this region on a different color scale in the inset. Areas not covered by MUSE observations are shaded in grey. {The white crosses mark the optical centers of the galaxies.} Our observations reveal a complex kinematical picture of the intra-group gas.}
\label{vel_map}
\end{figure*}

\begin{figure*}
\centering
\includegraphics[width=0.90\textwidth]{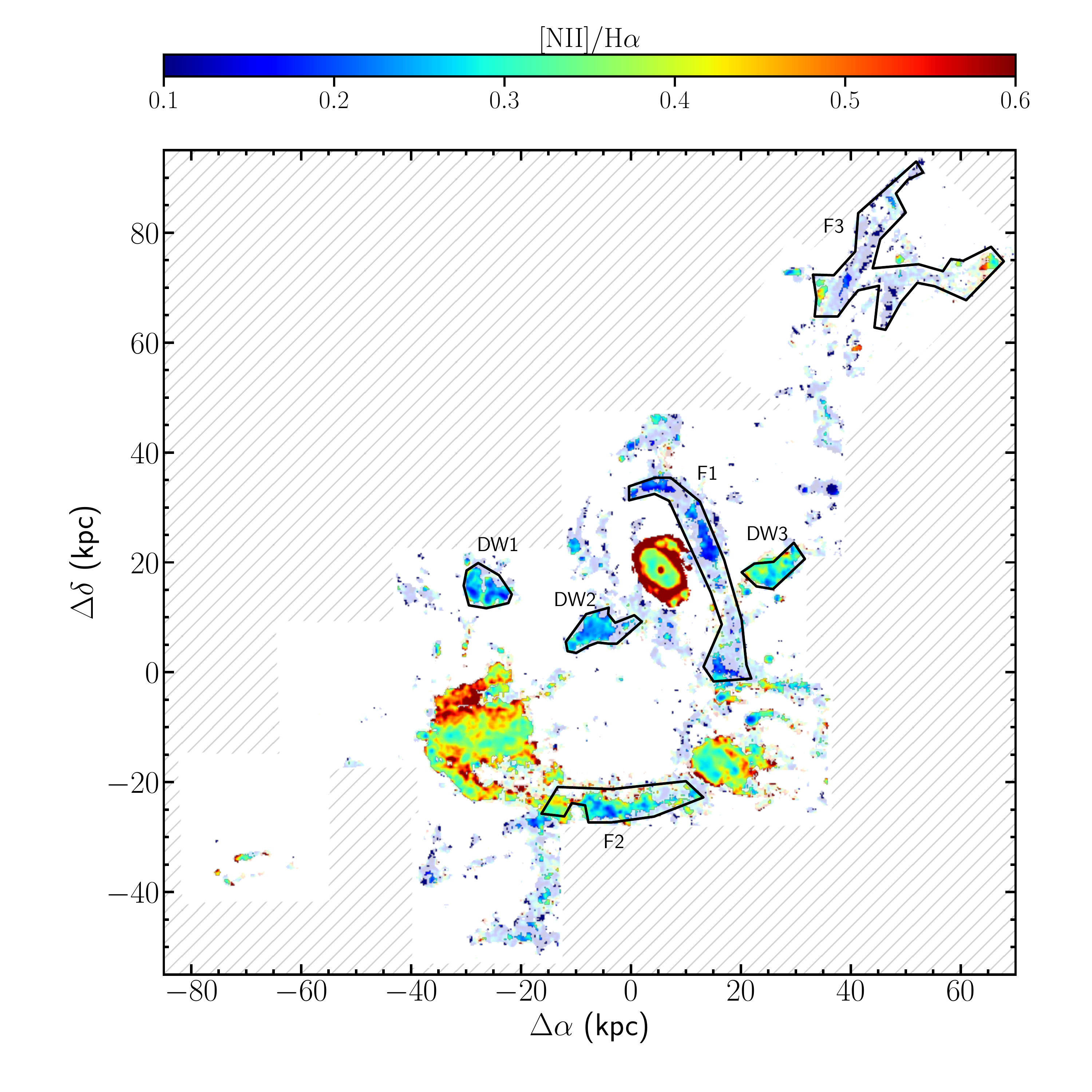}
\caption{Map of the [NII]/H$\alpha$ line ratio from pixels where the H$\alpha$ line is detected with $S/N>5$. Pixels where the $S/N$ of the [NII] line is less then 5 are shaded to reflect the higher uncertainty associated to the line ratio measurement. The black polygons identify the regions from which co-added spectra are extracted.}
\label{nii_ha_map}
\end{figure*}

Figure \ref{vel_map} shows the recessional velocity map from the fit of the H$\alpha$+[NII] line complex. Since the gas within \zgal120 has a much lower velocity compared to the rest of the map, we show a zoomed version of this region in the inset. This map reveals complex gas kinematics within the BIG group. First, the gas associated with \zgal125 does not show ordered rotation. Two streams of gas with velocities separated by $\simeq 250$ \kms\ can be observed over the stellar continuum of this object. The gas ionized in and around \zgal114 shows the highest velocities, which are also found in stripped debris in the North side of this galaxy {(at coordinates 25, -5 kpc in Figure \ref{vel_map})}. The gas extending West of \zgal125 shows a significant velocity gradient which cannot be fully traced back to the velocity of the filaments {(F1 in Figure \ref{nii_ha_map})} surrounding \zgal120. Moreover these filaments do not show signs of shear, at least at the spatial resolution of our MUSE data, which in turn makes a direct interaction of \zgal120 with the high velocity gas less likely. The dwarf star forming objects DW1, DW2, and DW3 exhibit velocities similar to \zgal125 and therefore can have been originated from tidal forces caused by the merging systems from which \zgal125 originated. Further evidences that this system is a recent post-merger remnant can be seen from the stellar velocity map.

\begin{figure*}
\centering
\includegraphics[width=0.90\textwidth]{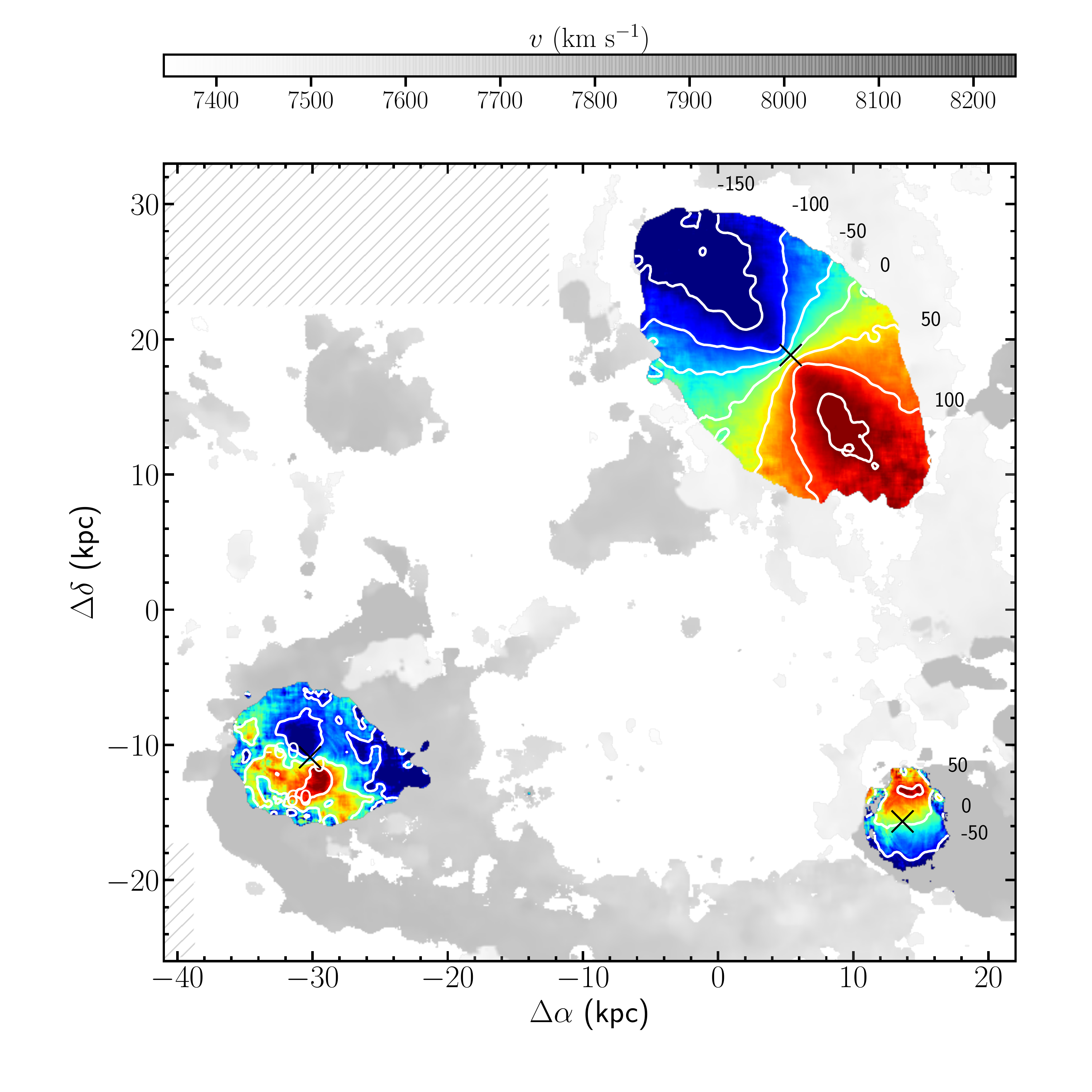}
\caption{Systemic velocity maps of stellar continuum for \zgal120, \zgal114, and \zgal125 from full spectral fitting of their stellar continuum spectra. The white contours mark the shape of iso-velocity lines. {The black crosses mark the optical centers of the galaxies.} The grey shaded background shows the extension and velocity of the ionized gas from Figure \ref{vel_map}. The massive galaxy \zgal120 has an extended rotation-dominated disk; on the contrary, the velocity map of \zgal125 is more chaotic and consistent with a recent merger event.}
\label{stellar_vel_map}
\end{figure*}

Figure \ref{stellar_vel_map} shows the map of the systemic velocity of the stellar component for 
\zgal120, \zgal114, and \zgal125, obtained from full spectral fitting of their stellar continuum spectra, as described in Section \ref{sec_fitting}. First we note the striking difference between the rotation maps of \zgal120 and \zgal125. The former shows ordered rotation up to twice the radius covered by the ionized gas. This is indicative of a minor or negligible role of gravitational interactions acting on this galaxy, which would have otherwise perturbed its stellar disk. On the other hand, the latter exhibits a complex stellar kinematic map. The peaks of the rotation curve are roughly aligned N-S, but {elsewhere the iso-velocity contours are mostly aligned E-W}. This shape further supports the evidence of a recent merging event occurred for this galaxy. Lastly, \zgal114 shows some form of ordered rotation. Due to its lower stellar mass, the dynamic range of the rotation curve is only $\pm 50$ \kms, but we find no sign of dynamical perturbations on the stellar disk based on current observations.   

\begin{figure*}
\centering
\includegraphics[width=0.90\textwidth]{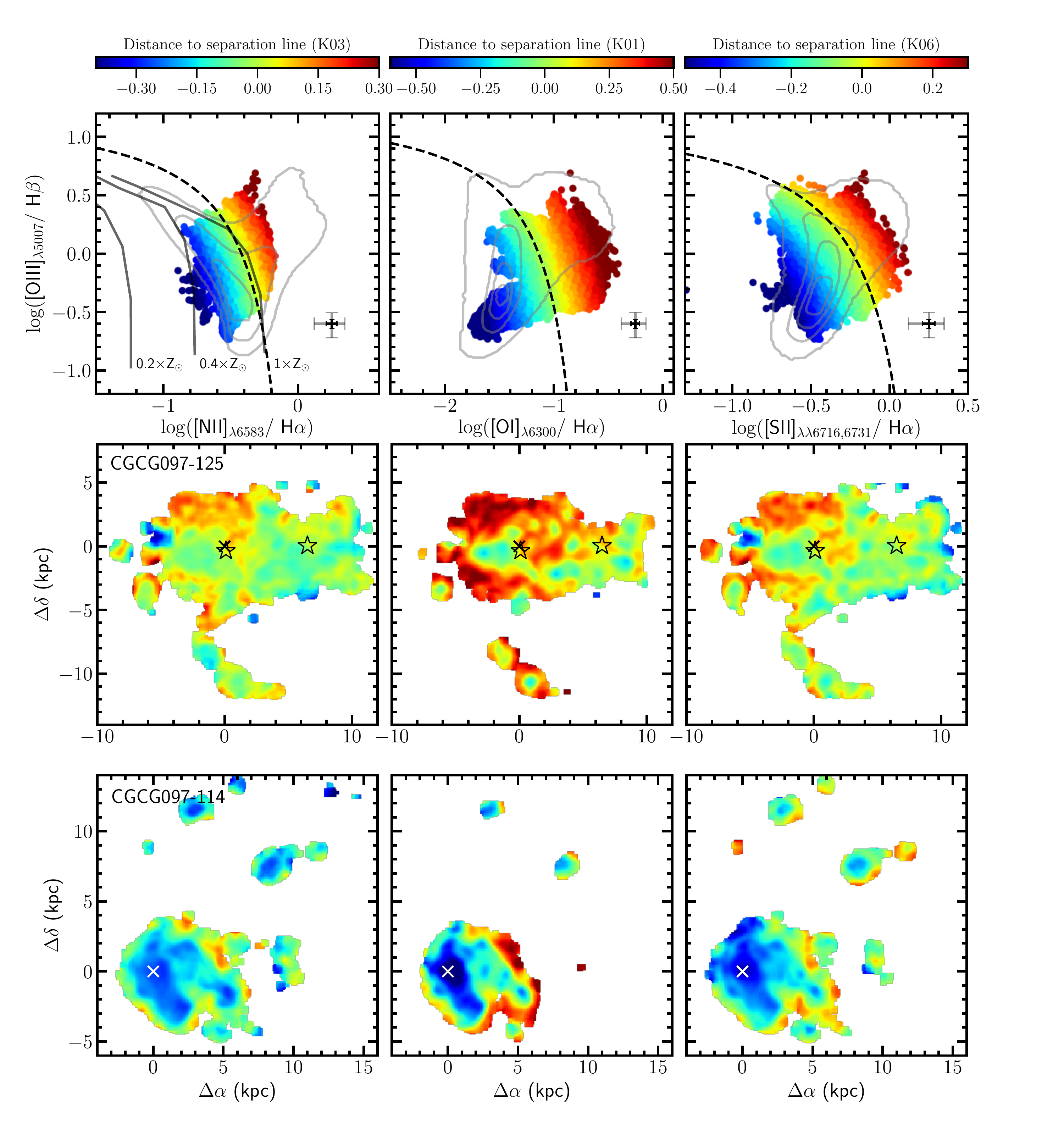}
\caption{Top row: BPT diagnostic diagrams [OIII]/H$\beta$ vs. [NII]/H$\alpha$ (Left Column), [OI]/H$\alpha$ (Middle Column), [SII]/H$\alpha$ (Right Column) derived from cutouts of the MUSE mosaic centered on \zgal125 and \zgal114. Only spaxels with a $S/N>5$ for all emission lines considered in each diagram are shown. The points are colour coded according to their distance from the dashed lines that separate stellar photoionization {(left of the line)} and AGN/shock ionization {(right of the line)} and are defined in \citet{Kauffmann03} (K03, left), in \citet{Kewley01} (K01, center), and \citet{Kewley06} (K06, right). The black and gray crosses indicate the typical error of the ratio of lines with an $S/N = 15$ and $S/N = 5$, respectively. The grey contours are obtained from the nuclear spectra of a random sample of 50,000 SDSS galaxies at $0.03 < z < 0.08$. The solid lines in the upper left panel show the tracks from photo-ionisation models of \citet{Kewley01} for three different metallicities $(0.2, 0.4, 1 Z_\odot)$. Middle and Bottom rows: The spatial (on-sky) position of the spaxels color coded as in the the top row panels is shown for CGCG097-125 and CGCG097-114 in the middle and bottom row panels, respectively. The open black stars mark the position of unresolved X-Ray sources. {The black crosses (white in the bottom panels) mark the optical centers of the galaxies.}}
\label{Galaxies_BPT}
\end{figure*}

\section{Emission line diagnostics and gas excitation} \label{emi_diagnostics}
The exquisite sensitivity of MUSE allows not only a robust determination of the emission line kinematics from the bright H$\alpha$+[NII] complex, but also an in-depth view of the physical conditions of the ionized gas from diagnostics involving weaker emission lines. Radiation fields arising from photo-ionization from hot stars or from harder sources, like Active Galactic Nuclei (AGN) or supersonic shocks produce different ratios in the intensity of different emission lines. As in other papers of this series \citep{Fossati16, Consolandi17}, and in other literature studies \citep{Merluzzi13, Poggianti17, Boselli18b, Boselli18c, Poggianti18}, we make use of three optical diagnostic diagrams \citep[BPT diagrams hereafter;][]{Baldwin81} based on the [OIII]$\lambda5007$/H$\beta$ ratio in combination with either [NII]$\lambda6584$/H$\alpha$, or [OI]$\lambda6300$/H$\alpha$, or [SII]$\lambda\lambda6716,6731$/H$\alpha$. 

In Figure \ref{nii_ha_map} we show the map of the [NII]/H$\alpha$ line ratio in the entire region mapped with MUSE. These lines are usually the strongest in our spectra, therefore this ratio gives us the best spaxel by spaxel picture of the gas ionization conditions and metallicity in the full field. The colored shaded regions in the map show the spaxels where the $S/N$ of the [NII] line is less than 5 and therefore the line ratios should be interpreted with more caution. Despite these caveats we can observe a large variety in the values of this ratio. We recall that for gas with a solar metallicity, [NII]/H$\alpha$ is around 0.3 for photoionization and higher in presence of shocks or AGN. In the disk of the three main galaxies (\zgal114, \zgal120, and \zgal125) we find significant spatial variations in the line ratio, especially in a ring like structure at the edge of the ionized gas region in CGCG097-120. This elevated values of the [NII]/H$\alpha$ ratio could be caused by the on-going stripping of gas at that radius.

\subsection{The gas excitation in \zgal114 and \zgal125}

In the galaxy disks, we detect at sufficient $S/N$ all the lines required to compile the BPT diagrams. We now explore in more details the full suite of line diagnostics in CGCG097-114 and CGCG097-125, while we describe the observational evidences of turbulent gas stripping from the disk of CGCG097-120 in Section \ref{obs_z120}.

The top panels of Figure \ref{Galaxies_BPT} show the three BPT diagnostic diagrams described above for individual spaxels where the lines have a $S/N>5$. The points are color coded by their distance from the line discriminating between photoionization and AGN/shock heating of the gas. Positive values are on the right of the separation lines, which are obtained from \citet{Kauffmann03}, \citet{Kewley01}, and \citet{Kewley06} for the [NII], [OI], and [SII] BPT diagrams respectively. The black solid lines in the upper left panel show the effects of gas metallicity on the line ratios. Metallicity affects more significantly the [NII]/H$\alpha$ ratio, as opposed to the [OIII]/H$\beta$, and the best separation of the iso-metallicity curves occurs at low [OIII]/H$\beta$ ratios. The middle and bottom panels of Figure \ref{Galaxies_BPT} show the spatial distribution of the points (spaxels) shown in the upper panels, for the area around \zgal125 and \zgal114 respectively. Although the ratios in the BPT diagrams are chosen to minimize the effects of dust extinction, by taking nebular lines close in wavelength, the [OI]/H$\alpha$ is the ratio that could be more effected by dust reddening. Using the ratio H$\alpha$/H$\beta$, extracted on an integrated aperture over the galaxy bodies, we find that the attenuation of the nebular lines due to dust is $A_V = 1.30 \pm 0.09, 1.13 \pm 0.12$ for \zgal125 and \zgal114 respectively. Assuming these values and the \citet{Calzetti00} reddening curve, the [OI]/H$\alpha$ ratio would be underestimated by $\simeq4-5\%$, not affecting our results. 

We immediately witness a different behaviour of the ionized gas excitation in the bodies of the two galaxies. 
For \zgal114, and its stripped star forming knots on the N side {(at coordinates 3,12 kpc and 9,7 kpc in the bottom panels of Figure \ref{Galaxies_BPT})}, we observe the lowest values of the line ratios, consistent with ionization from on-going star formation, possibly at sub-solar metallicity as one could expect from the low stellar mass of this galaxy \citep{Tremonti04}. In the ram pressure stripped tail downstream of \zgal114,  line ratios appear inconsistent with photoionization and are likely dominated by shocks and turbulence in the stripped gas. The transition appears sharper in the BPT diagram using the [OI]/H$\alpha$ ratio, which has been generally regarded as the most sensitive in these conditions \citep{Yoshida08, Fossati16, Poggianti18}. However we note consistency in all three BPT diagrams used in this work. 

Conversely, the BPT diagrams for \zgal125 show elevated line ratios almost everywhere in the galaxy disk. Only the inner part of the stellar disk exhibit line ratios possibly consistent with photoionized gas at solar metallicity, in the [NII]/H$\alpha$ BPT diagram. When the other line ratios (which are less sensitive to metallicity) are used, we observe a high incidence of spaxels dominated by shocks. This is especially true in the E side of the galaxy, where high velocity gas is observed in Figure \ref{vel_map}. These elevated line ratios are another indication of the recent merger event which \zgal125 has undergone. Indeed the BPT diagrams for this galaxy closely resemble the ones observed using the CALIFA integral field spectrograph in the Mice (NGC4676) merger \citep{Wild14}.

While in both galaxies we find no nuclear enhancement of the optical line ratios related to the presence of AGN, we also analyzed the archival {\it Chandra} X-Ray observations of this region. Particularly, we analyzed seven observations (ObsID: 17199, 17201, 17589, 17590, 17591, 17592, 18755) from the {\it Chandra} archive where both galaxies are within 4 arcmin of the optical axis. Standard {\it Chandra} analysis has been done. The combined clean time is 269 ks with ACIS-I. There is an obscured X-ray unresolved source at the position of \zgal125's nucleus (R.A. 11:44:54.84 Dec. +19:46:34.5). The intrinsic absorption is quite high, 1.9$\pm0.3 \times 10^{25}$ cm$^{-2}$ and the 2 - 10 keV unabsorbed luminosity is 3.1$\times10^{41}$ erg s$^{-1}$ (for comparison, the 2 - 10 keV unabsorbed luminosity for \zgal120's nucleus, which also hosts an AGN, is 4.2$\times10^{40}$ erg s$^{-1}$). Such a high intrinsic absorption and relatively low X-ray luminosity probably explain the lack of optical nuclear lines. Interestingly, there is another X-ray unresolved source at R.A. 11:44:53.81, Dec. +19:46:35.4, close to the other optical component in \zgal125. There is no evidence for intrinsic absorption and the 2 - 10 keV unabsorbed luminosity of this source is 5.5$\times10^{39}$ erg s$^{-1}$, if in A1367.
Their positions (shown as open black stars in the middle row panels of Figure \ref{Galaxies_BPT}) roughly correspond to the centers of the two distinct kinematic components of the ionized gas {see Figure \ref{vel_map})}, further reinforcing the merger scenario.
For \zgal114, indeed its nucleus is not detected in hard X-rays (2 - 8 keV).

\subsection{A face-on stripping example in CGCG097-120} \label{obs_z120}
We now investigate the morphology of the ionized gas in the massive galaxy \zgal120. It is immediately clear from Figure \ref{Halpha_map} that the ionized gas extends only 5.5 kpc from the galaxy nucleus along the major axis. The stellar disk, instead, is at least twice as extended. This abrupt truncation of the gaseous disk is one of the strongest indications of ram pressure stripping, and typically occurs in all the gas phases (atomic, molecular, ionized) and in the dust distribution \citep{Koopmann01, Boselli06, Fumagalli08, Fumagalli09, Cortese10, Fossati13, Boselli14a}.

\begin{figure*}
\centering
\includegraphics[width=0.90\textwidth]{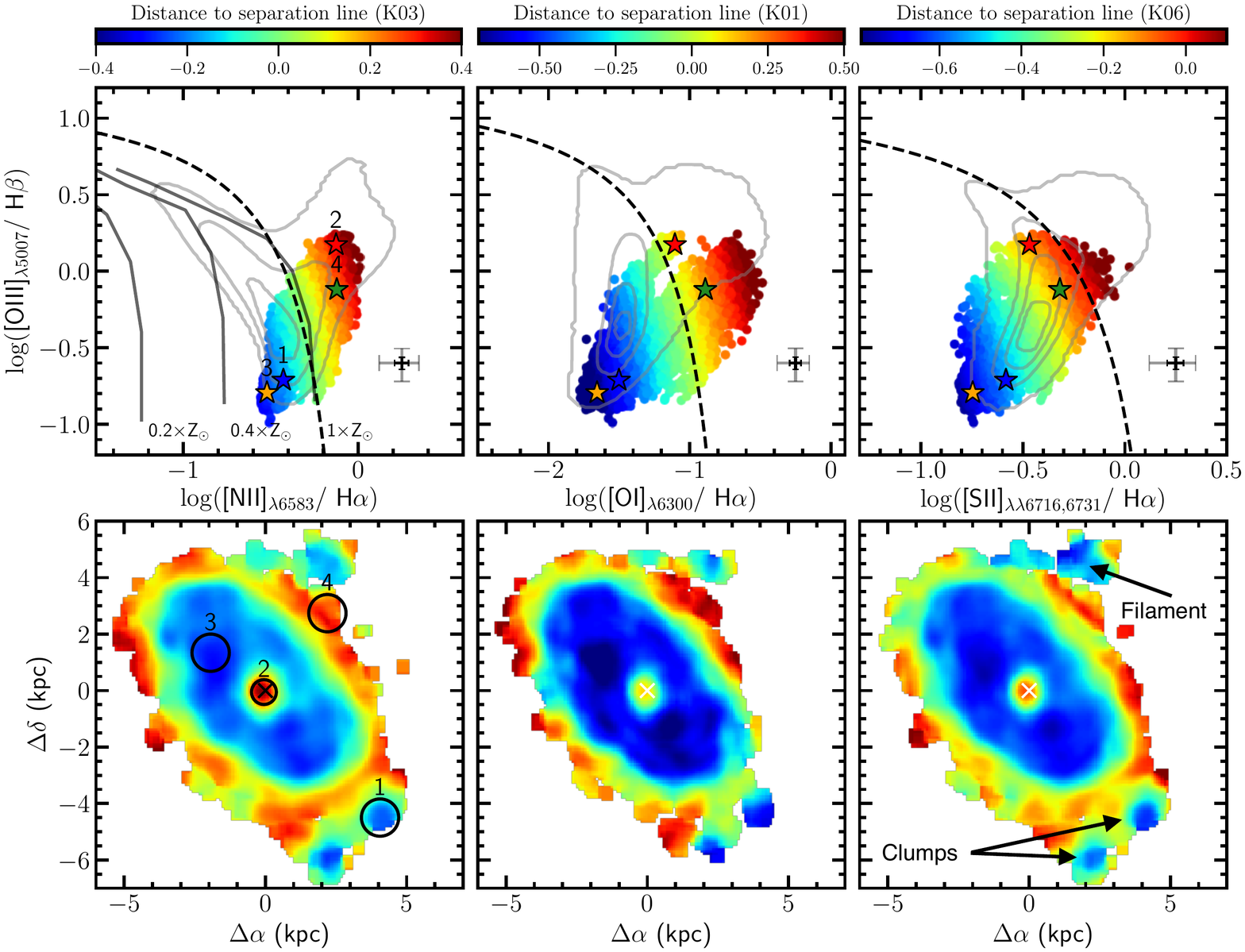}
\caption{Same as Figure \ref{Galaxies_BPT} but for \zgal120. Four characteristic regions are highlighted by numbered black circles on the bottom left panel, their composite spectra are shown in Figure \ref{templ_spec_120}, and the line ratios are shown as big stars in the top panels. The number above the star in the top left panel denotes the region. {Two star forming clumps and a filament described in the text are labelled in the bottom right panel.}}
\label{CGCG120_BPT}
\end{figure*}

\begin{figure*}
\centering
\includegraphics[width=0.95\textwidth]{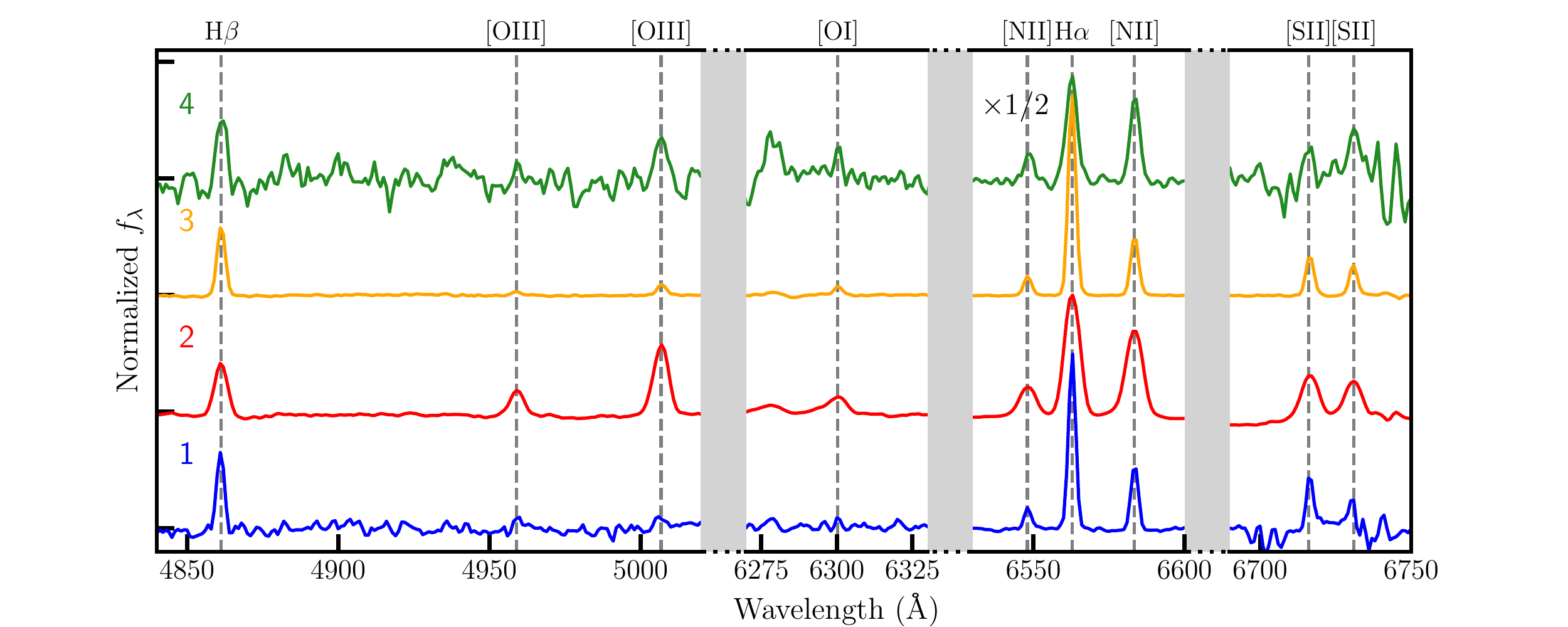}
\caption{Co-added rest-frame spectra of the four regions defined in the emission line region of CGCG097-120 from Figure \ref{CGCG120_BPT}. The spectra are normalized such that the flux of the H$\beta$ line is 1 and shifted along the vertical axis to avoid overlap. The H$\alpha$+[NII] spectral cutout has been scaled by a factor 1/2 compared to the other spectral regions. Only the emission lines relevant to the BPT diagnostic diagrams are shown.}
\label{templ_spec_120}
\end{figure*}

In Figure \ref{CGCG120_BPT}, we show the BPT diagnostic diagrams presented in Figure \ref{Galaxies_BPT}, but for \zgal120. Also in this case we find that dust extinction does not affect our results. We measure an average attenuation of $A_V = 2.06 \pm 0.11$ over the region with emission lines, which turns into an underestimate of the [OI]/H$\alpha$ ratio of at most 8\%. The spatial maps are rich of features, which we interpret with the help of co-added spectra from four circular apertures placed on regions of particular interest: 1) A compact knot of star formation at the edge of the ionized gas disk; 2) the nuclear region; 3) the galaxy star forming disk; and 4) the edge of the line emitting region. The co-added spectra have been extracted from the unsmoothed datacube by summing the individual spectra, and are shown in Figure \ref{templ_spec_120}. 
We then fit the emission lines as described in Section \ref{obs_datared} for each of the co-added spectra and we show the line ratios as big stars in Figure \ref{CGCG120_BPT} following the color coding of Figure \ref{templ_spec_120}. 

\begin{figure}
\centering
\includegraphics[width=1.00\columnwidth]{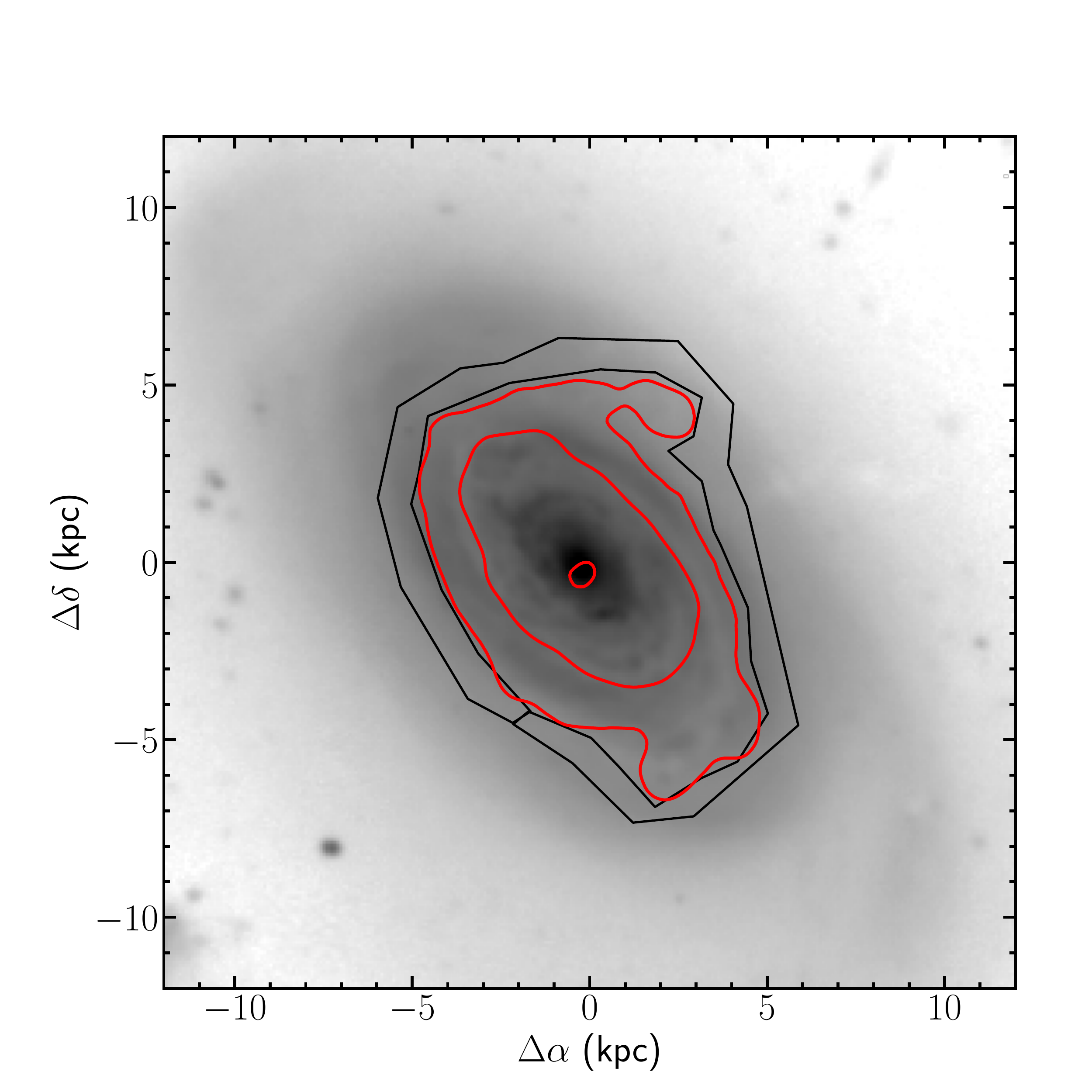}
\caption{Subaru $B-$band image of \zgal120. The red contours show the extent of the H$\alpha$ emission at surface brightness levels of $10^{-15}$, $10^{-16}$, and $10^{-17}$ $\rm{erg~s^{-1}~cm^{-2}~arcsec^{-2}}$. The area within the black polygon is the region from which we extract multiwavelength photometry and MUSE spectroscopy for the stellar population analysis.}
\label{map_region_120}
\end{figure}

In the galaxy nucleus we find elevated line ratios, consistent with the presence of an active galactic nucleus. However [OI]/H$\alpha$ does not reach its maximum value in this region, but rather in the outer ring, which we interpret as an indication that AGN ionization does not enhance [OI] as much as other phenomena (e.g. shocks) do. 
The galaxy disk is instead characterised by lower line ratios, well within the photoionization region of the BPT diagrams, and a cold kinematics ($\sigma \simeq 27$ \kms\ as opposed to $\sigma \simeq 108$ \kms\ in the galaxy nucleus), typical of star forming disks in the local Universe \citep{Leroy08, Green14}.

The outer ring of ionized gas shows again elevated line ratios, reaching the highest values in the map for the [OI]/H$\alpha$ ratio. These values are in the AGN/shock region of the BPT diagrams (with the exception of the [SII] BPT), but with lower [OIII]/H$\beta$ ratios compared to the AGN emission from the nucleus. The gas kinematics is also turbulent with a velocity dispersion of $\simeq 60$ \kms. Shocks from the on-going stripping process due to the high velocity motion of \zgal120 in the cluster potential are likely to produce these features, including a strong correlation between shocks and high gas velocity dispersion as found by \citet{Rich11, Ho14, Fossati16, Boselli18b}. 

Lastly, we find two clumps of ionized gas in the S-W side of the disk, and a filament that resembles a spiral arm in the N-W direction which are located outside the high line ratios ring. These regions stand out for lower line ratios, which appear very similar to the galaxy inner star forming disk. They are likely to host molecular clouds which are too dense to be directly stripped by ram pressure \citep{Tonnesen09}. As a result, they become decoupled from the rest of the ISM, and remain star forming in a mostly stripped region until they run out of gas for further star formation, as proposed by \citet{Abramson14}. 

While the emission lines provide us with diagnostics on the on-going effects of ram pressure stripping, a complete picture of the past history 
of \zgal120 and its role in the BIG group can only be obtained by analysing the stellar populations outside the stripping radius. Figure \ref{map_region_120} shows the Subaru $B-$band image of \zgal120. The red contours outline the line emitting region, while the black polygon shows the stripped region where we aim to reconstruct the past star formation history.  To do so we follow the approach introduced in \citet{Fossati18}, and we refer the reader to that work for the details of the procedure. 

Briefly, we extract a co-added MUSE spectrum in the region of interest, using the method outlined above and using the stellar velocity map to remove the signatures of the disk rotation prior to co-adding the pixels. Then we extract aperture photometry in the same region in 8 photometric bands. We use Galaxy Evolution Explorer \citep[GALEX, ][]{Martin05} data in the FUV and NUV bands from program GI5-063 (PI Sakai); Subaru $B-$ and $R-$ band images from \citet{Yagi17}; and {\it Spitzer} \citep{Werner04} images taken with the IRAC instrument at 3.6, 4.5, 5.8, and 8.0 $\mu$m from \citep{Fazio04}. These bands offer coverage of young stellar populations (FUV to $B$), the bulk of the stellar mass (IRAC 3.6 $\mu$m) and the dust reprocessed emission (IRAC 5.8 and 8.0 $\mu$m). They are therefore of invaluable importance in breaking the degeneracies between the mixing of stellar populations of different ages and the dust absorption. 

In order to reconstruct the quenching history of \zgal120 we need to model its unperturbed star formation history (SFH) at the radius of the truncation ($\simeq 5-6$ kpc). Following \citet{Fossati18} we use the multizone models for the chemical and spectro-photometric unperturbed disk evolution from \citet{Boissier00} and \citet{Munoz-Mateos11} from which we extract the SFH for a galaxy of rotational velocity similar to that of \zgal120 (150 \kms) and at the observed truncation radius. We modify the unperturbed SFHs by introducing an exponential decline in the SFR to parametrise the quenching event. The choice of an exponential quenching pattern allows flexibility capturing models that range from an almost instantaneous event, to a more marginal and gradual suppression of the star formation activity. The fitting function has therefore two parameters: $Q_{\rm Age}$ which is the look-back time for the start of the quenching event, and $\tau_{\rm Q}$ that is the exponential timescale of the quenching. To construct the model spectra we feed the grid of SFHs into \citet{Bruzual03} high-resolution models assuming a stellar solar metallicity, and we further include nebular lines and dust reprocessing.

\begin{figure*}
\centering
\includegraphics[width=0.98\textwidth]{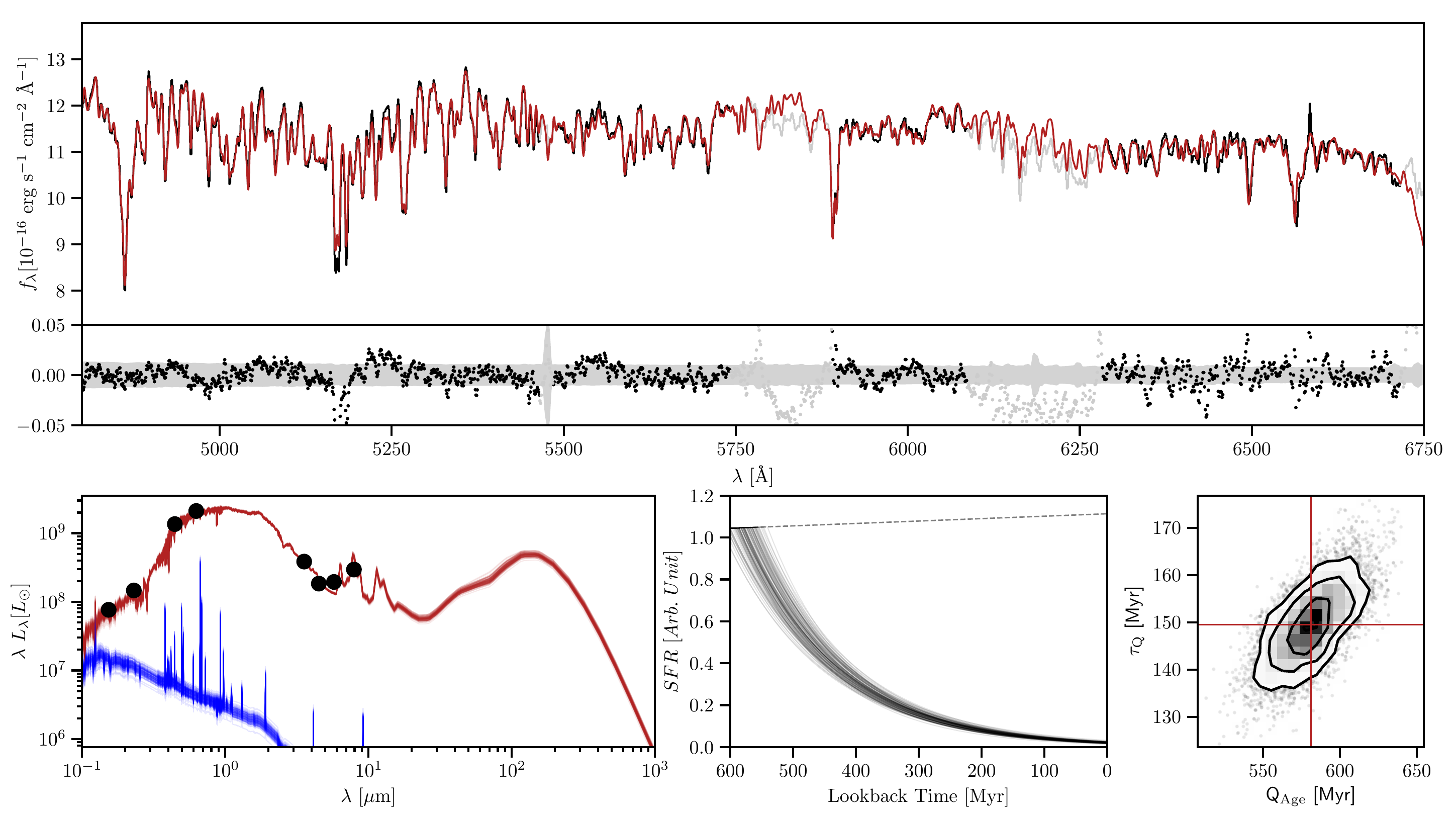}
\caption{Results of the MC-SPF fitting for \zgal120. Upper panel: MUSE spectrum (black) and best fit model (dark red). Regions where the spectrum is plotted in grey are not used in the fit, and they correspond to bright skyline groups. The fit residuals (Data - Model) are shown below the spectrum and the grey shaded area shows the 1$\sigma$ uncertainties. Lower left panel: Photometric data points in black. The dark red lines are the total model including the dust emission, while the blue lines show the contribution to the stellar continuum and nebular line emission from young stellar populations (Age < 10 Myr). Different lines are obtained by randomly sampling the posterior distribution. Lower middle panel: Reconstructed SFH from the fitting procedure. The grey dashed line shows the evolution of the unperturbed SFH. Lower right panel: Marginalised likelihood maps for the $Q_{\rm Age}$ and $\tau_{\rm Q}$ fit parameters. The red lines show the median value for each parameter, while the black contours show the 1, 2, and 3 $\sigma$ confidence intervals.}
\label{SPSfitting}
\end{figure*}

We fit the data to the models using the Monte Carlo Spectro-Photometric Fitter \citep[MC-SPF;][]{Fossati18}, a python code that jointly fits photometry and spectra to parametric models. MC-SPF uses the {\sc MultiNest} \citep{Feroz08, Feroz09, Feroz13} code to sample the posterior distribution and derive best fit parameters and reliable uncertainties in a Bayesian framework. 
Figure \ref{SPSfitting} shows the MC-SPF output.
The top panel shows the co-added MUSE spectrum with the best fit model (red line) superimposed, the bottom left panel shows the best fit spectrum (dark red line) and the photometric data points in black. The bottom middle panel shows the reconstructed SFH during the quenching phase and the amount of attenuation of the SFR compared to the unperturbed models, which is found to be consistent with unity, i.e. complete suppression of star formation. The bottom right panel shows the best fit values for $Q_{\rm Age}$ and $\tau_{\rm Q}$, and their associated confidence intervals. 

The joint fit of spectroscopy and photometry reveals a quenching event that, for the region at $5-6$ kpc from the center,  started $\simeq 570$ Myr ago and proceeded gently with an exponential e-folding time of $\simeq150$ Myr. Moreover the best fit model shows that dust is not fully stripped at this radius ($A_V = 0.33\pm0.02$), which indicates that the stripping event has been more gentle than the nearly instantaneous stripping found for instance in NGC4330 within the Virgo cluster \citep{Fossati18}. Based on this analysis we find that \zgal120 is undergoing a ram-pressure stripping event that has caused a truncation of the gaseous disk and the suppression of star formation at $r>5$ kpc, finding instead no perturbation on the stellar disk kinematics. The stripping causes an increase in BPT line ratios and in the gas velocity dispersion in a narrow ring at the edge of the star forming disk, probably due to shocks. In the discussion Section we will come back to the implications of these results on the role of \zgal120 in the BIG group.

\begin{figure*}
\centering
\includegraphics[width=0.95\textwidth]{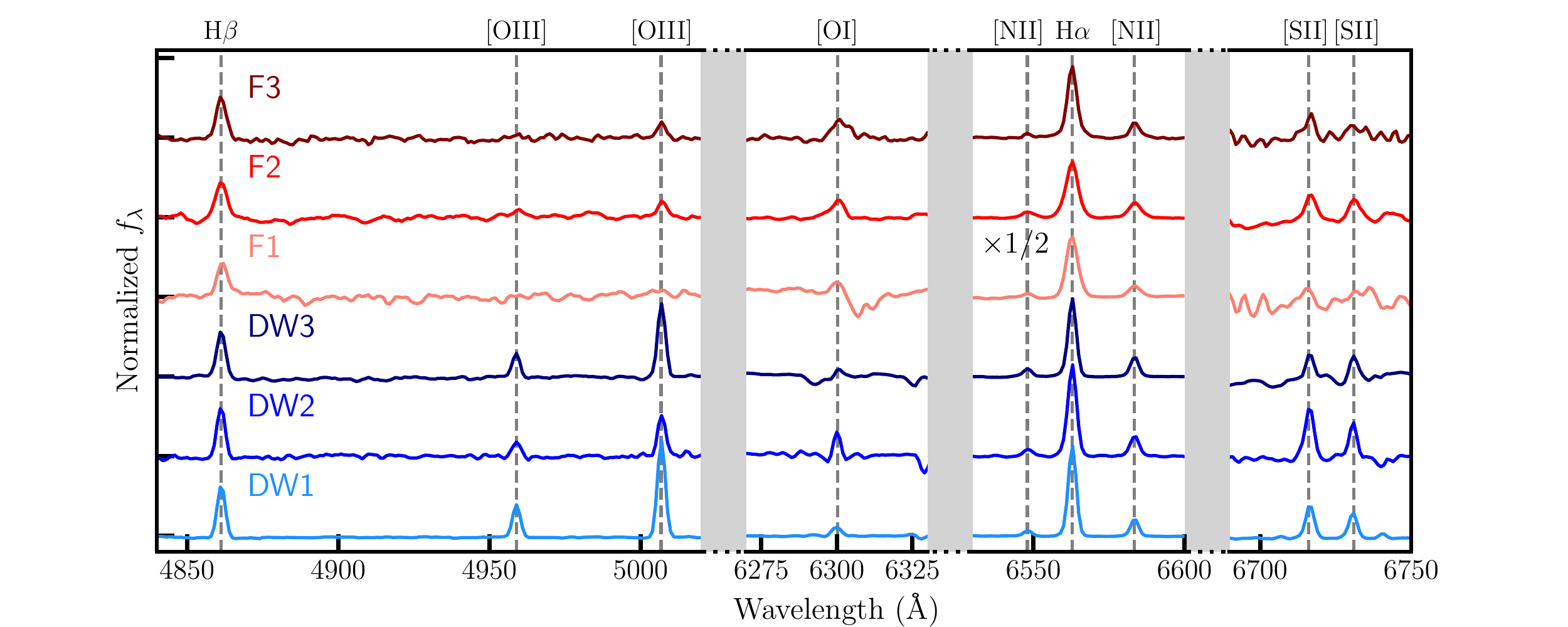}
\caption{Same as Figure \ref{templ_spec_120} but for the Co-added rest-frame spectra of the three dwarf emission line systems and the three diffuse gas filaments defined in Figure \ref{nii_ha_map}. High surface brightness knots, detected on the H$\alpha$ map (see the text for details) have been excluded from the filaments spectra.}
\label{templ_spec_tails}
\end{figure*}

\begin{figure*}
\centering
\includegraphics[width=1.00\textwidth]{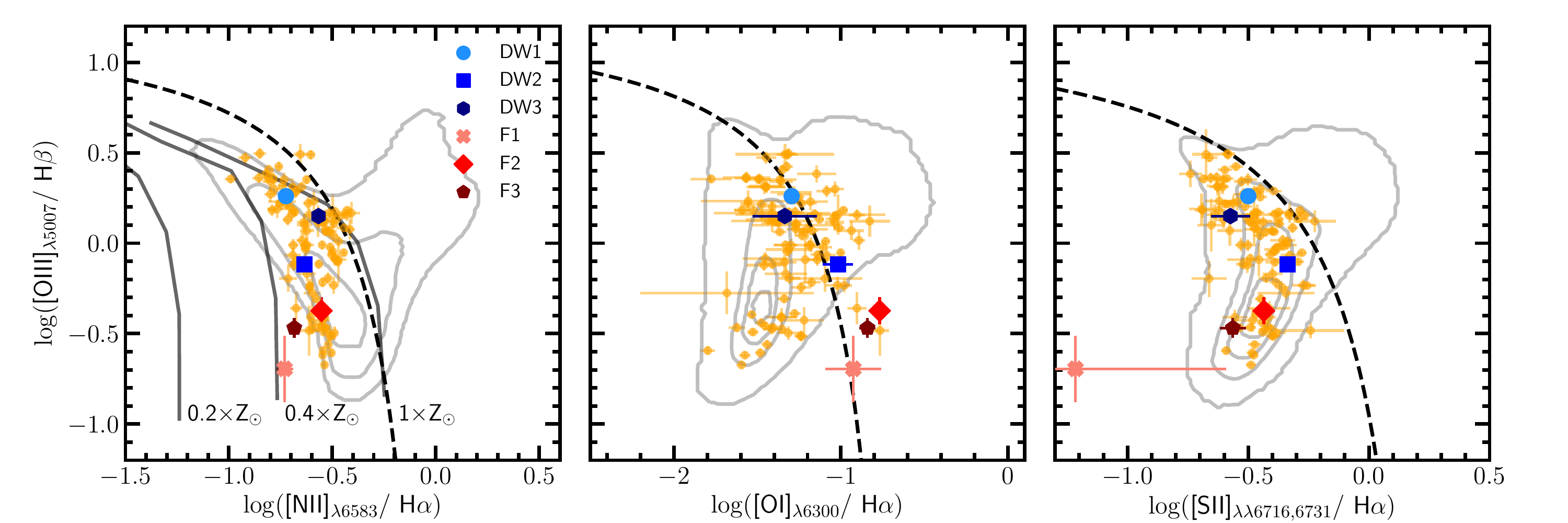}
\caption{BPT diagnostic diagrams [OIII]/H$\beta$ vs. [NII]/H$\alpha$ (Left Panel), [OI]/H$\alpha$ (Middle Panel), [SII]/H$\alpha$ (Right Panel) for the composite spectra extracted from the six characteristic regions of the BIG group shown in Figure \ref{templ_spec_tails} and for individual compact knots (small orange circles). The error bars are obtained with a MonteCarlo error propagation of the variance of individual spaxels. The grey contours, and solid and dashed black lines are as in Figure \ref{Galaxies_BPT}. }
\label{BPT_1D}
\end{figure*}

\subsection{The physical properties of the diffuse gas filaments and the dwarf systems}

We now turn our attention to the tendrils of diffuse gas which appear to connect the galaxies and the dwarf systems labelled in Figure \ref{Halpha_map}. Due to the very low surface brightness of the emission lines required to build the BPT diagrams we co-add spaxels from selected spatial regions to extract a 1D spectrum with high $S/N$. We identified three main filaments (labelled in Figure \ref{nii_ha_map}), one surrounding \zgal120 (F1), one extending between \zgal114 and \zgal125 (F2), and one belonging to the inner part of the NW long tail of ionized gas (F3). We also extract the average spectrum of the dwarf systems (DW1 to DW3), aiming to better understand their origin.

Since our goal is to investigate the emission line properties of the faint and diffuse gas in the filaments we have to remove high surface brightness compact clumps (knots hereafter), for which we study the gas properties independently. Following \citet{Poggianti17} we detect bright knots as local minima in the Laplace filtered H$\alpha$ image. This filtering technique highlights the flux peaks embedded in a smoother distribution. Due to the high surface brightness of the knots we apply the Laplace filter to the unsmoothed (seeing limited) H$\alpha$ image, to avoid the blending of different knots. A total of 111 local minima are identified in the filtered image and these are also confirmed with visual inspection. We extract the spectra of these regions using a circular aperture with 1.0 arcsec radius, as described in the previous Section, and we show the line ratios on the BPT diagrams in Figure \ref{BPT_1D}. The uncertainties on the line ratios are obtained via a MonteCarlo perturbation of the individual values in the 3D datacube, using a Gaussian noise with $\sigma$ equal to the standard deviation of the combined cubes derived by our data reduction pipeline. The spectra are then extracted on 100 realisations of these perturbed cubes and the distribution of the emission line fluxes from the fit determines the final uncertainty.

The distribution of the bright knots in the BPT diagrams follows the sequence of HII regions, i.e. the sites of on-going star formation and photoionization of the gas by young stars. Indeed, using the [NII] BPT we find 102/111 knots in the region of the diagram dominated by photoionization, and an even slightly higher fraction (107/111) is found if we use the [SII] BPT. On the other hand, this fraction drops to $\simeq70\%$ (79/111) if we use the [OI] BPT. This small tension could originate from the fact that we observe the knots embedded in diffuse gas which is likely shock heated and the [OI] emission could originate from this diffuse component which is super-imposed to the HII spectra along the line of sight. Indeed, being [OI] the transition that is most sensitive to shocks and other complex excitation mechanisms, this behaviour is not entirely unexpected as first found by \citet{Fossati16} and later confirmed with higher number statistics by \citet{Poggianti18}. Lastly we note that the HII regions span $\simeq1$ dex in the values of the [OIII]/H$\beta$ ratio. This ratio is correlated to the ionization parameter \citep[the ratio of the rate of ionizing photons to the mean electron density, ][]{Osterbrock06}. From the ratio of the [SII]$\lambda\lambda6716,6731$ lines, we find that the electron density is $<100~{\rm cm^{-3}}$ and often reaches the low density limit where the line ratio is insensitive to a change in density. These values are consistent with the work by \citet{Poggianti18}, who found an average density of $50~{\rm cm^{-3}}$ for a large sample of HII regions in ram pressure stripped galaxies. Assuming that the density is not the main driver of the range in ionization parameter, the other variable is the rate of ionizing photons, which in turn depends on the age of the star forming region. We are therefore observing HII regions at different (but still young, $<50$ Myr) ages, which are born {\it in-situ} in the filaments of diffuse gas or in the disk of the galaxies in the BIG group. 

After the identification and characterisation of the compact knots, we now turn to the diffuse gas filaments and the dwarf star forming systems. To extract their co-added spectra we follow the procedure presented in \citet{Fossati16}. We remove the compact knots from the filaments (F1 to F3) and select the spaxels within the masks shown in Figure \ref{nii_ha_map} from the un-smoothed datacube. We further subtract the stellar continuum of \zgal120 using {\sc GANDALF}, which would otherwise dominate the signal for F2. We then bring each spectrum to rest-frame by using the velocity field derived from our {\sc kubeviz} fit and we further correct for the continuum shape of each spectrum by subtracting a $4^{\rm th}$ degree polynomial fit to the spectral continuum (masking emission lines). The spectra are then summed, their emission lines are fit and we normalise them to the H$\beta$ flux. The uncertainties are obtained as described above for the knots. The co-added spectra of the filaments and dwarf systems are shown in Figure \ref{templ_spec_tails}.

The diffuse gas filaments show a complex picture of line ratios as shown in Figure \ref{BPT_1D}. They occupy the photoionization region of the [NII] BPT diagram, possibly with a subsolar metal abundance for F1 and F3. The filament F1 shows the lowest [OIII]/H$\beta$ ratio, which indicates a soft ionizing spectrum (the photon energy required to doubly ionize an oxygen atom is 35eV, much higher than the 13.6eV needed to ionize hydrogen). The filaments occupy the photoionization region also in the [SII] BPT, while they show elevated [OI]/H$\alpha$ ratios, consistent with shocks or thermal heating as found by \citet{Fossati16} and later confirmed by \citet{Poggianti18} using a larger sample of ram pressure stripped galaxies. Although in presence of physical processes more complex than stellar photoionization it is not possible to reliably measure the gas metallicity from the line ratios, we note that the filament F1 has the lowest ratio of metal to hydrogen lines in all the BPT diagnostic diagrams, an evidence that could reinforce the possibility that the gas in F1 has a different origin compared to the other filaments.  

The dwarf systems show instead the typical line ratios of star forming (HII) regions which include an elevated [OIII]/H$\beta$ and a low [OI]/H$\alpha$ ratio. Indeed, they occupy the HII regions position in the BPT diagrams in Figure \ref{BPT_1D}, with DW2 showing the lowest [OIII]/H$\beta$ ratio. The metallicity of the brightest knots in these systems has been studied by \citet{Cortese06} using long slit spectra that covered the [OII], [OIII], H$\beta$, H$\alpha$, and [NII] lines. These authors averaged the results of several metallicity indicators and found  the oxygen abundance to be 12+$\log(\rm{O/H})\simeq8.50-8.75$ for the three systems with some scatter among different knots in the same dwarf system. We re-evaluate the metallicity of the dwarfs in light of the full spatial coverage of MUSE and more modern calibrations \citep{Curti17} of the line ratios into oxygen abundance. However, we are limited to use the O3N2 indicator ([OIII]/H$\beta~\times~$H$\alpha$/[NII]) due to the lack of [OII]$\lambda3727$ in the MUSE spectral range. Using the O3N2 indicator we find 12+$\log(\rm{O/H})\simeq 8.53, 8.63, 8.59$, for the dwarf systems DW1 to DW3 respectively. Given the high $S/N$ of the co-added spectra, the dominant uncertainty on the metallicity measurements resides in the calibration of the line ratios into oxygen abundance. For the O3N2 indicator, \citet{Curti17} find this uncertainty to be $\simeq 0.1$ dex. With this uncertainties, our modern determination of the oxygen abundance is in good agreement with \citet{Cortese06}. Using the calibrations of \citet{Zibetti09} and the deep $B$ and $R$ band images taken with the Subaru telescope, we computed the stellar mass of the dwarf systems. These are $\log(M_*/{\rm M_\odot})\simeq7.37, 7.08, 6.91$ for DW1 to DW3 respectively. We caution the reader that these stellar masses have to be treated as upper limits, as the $B$ and $R$ broadband filters include flux from the emission lines. In this range of stellar mass we would expect the metallicity to be $\simeq$0.8-1.0 dex lower {than the one measured} if these systems were to follow the mass-metallicity relation \citep{Andrews13} for dwarf galaxies. Their metal enrichment thus supports the scenario proposed by \citet{Sakai02} and \citet{Cortese06} that these dwarf systems have originated from {multiple tidal stripping events} during the merger event of \zgal125.

Lastly, we note that, given the similar metallicity of all the dwarf systems, the lower [OIII]/H$\beta$ ratio shown by DW2 could be caused by an older age of the stellar populations in this system. However, this can only be confirmed with very deep observations allowing a direct measurement of the age of the stellar populations. 

\section{Discussion} \label{discussion}

The mosaic of $4\times 4$ arcmin of deep MUSE integral field observations centred on the BIG group reveals a very complex picture of ionized gas filaments, and perturbed galaxies where tidal and hydrodynamic interactions are likely at play. \citet{Cortese06} combined deep H$\alpha$ and broad-band imaging with long slit spectroscopy to decipher this complex puzzle. These authors concluded that \zgal125 is the result of a merging event of a relatively massive galaxy $(M_*\sim 10^{10} \rm{M_\odot}$) with a lower mass satellite, and that the orbits of these objects before the merger could explain the emission line trails. However, the role of \zgal120 and \zgal114, and the origin of the gas ionization remained unclear. In light of the new data we can now re-address these questions.

\subsection{Is the Blue Infalling Group made by \zgal125 and \zgal114?}

There is little doubt that \zgal125 is a post-merger object. First, the presence of resonance shells in the deep Subaru images {(light excess on the East side of \zgal125 in Figure \ref{BIG_RGB_ima})} is a clear indication of a gravitational perturbation of the potential field \citep{Duc11}. These features are produced by stars bouncing back towards the minimum in the potential well after reaching the maximum distance from it due to their kinetic energy. The fact that the shells extend more on the E side of \zgal125 could mean that one (or both) the merging galaxies were approaching from the W-NW direction {and therefore their stars have travelled to the E thanks to their initial kinetic energy before being pulled back towards the center of the potential}. The filament F2 (see Figure \ref{nii_ha_map}), which is kinematically connected to the gas in \zgal125 
is likely to be the tracer of this interaction. The MUSE data reinforced this scenario by showing a perturbed stellar velocity field, typical of post-merger remnants \citep{Barnes92}. It is also interesting that the stars are rotating roughly along an E-W axis while the gas shows signs of rotation along a N-S axis. 
Moreover, we find enhanced line ratios throughout the entire disk of \zgal125, which implies a turbulent ionized gas medium which still has to settle in the galaxy potential. This scenario is further complicated by the presence of the dwarf star forming systems {(DW1-DW3 in Figure \ref{nii_ha_map})} on the North West side of \zgal125. Their metallicity is 0.8-1.0 dex higher than what would be expected for objects of similar stellar mass, as such \citet{Sakai02} and \citet{Cortese06} have suggested they have formed from tidally disrupted material during the merging of \zgal125. Their position with respect to \zgal125 could be an indication of the trajectory of the pre-merger galaxies.
Indeed, the long gas tail which we have partially observed with MUSE (F3) is pointing in an opposite direction compared to the A1367 cluster center, which is highly suggestive of an infall event for \zgal125 and \zgal114 into the cluster, while the merger was taking place.

The role of \zgal114 remains unclear, despite the wealth of data provided by the MUSE observations. There is no doubt that the galaxy is undergoing ram pressure stripping, possibly by the hot halo of A1367. This is supported by the tail downstream of the stellar disk, and the presence of stripped HII regions in the N direction, exhibiting extended filaments pointing in the same direction as the main tail behind the galaxy. 

The main tail shows signs of shock heating (enhanced [OI]/H$\alpha$ ratio) which is a common feature of stripped tails \citep{Yoshida12, Fossati16, Consolandi17, Poggianti18, Boselli18b}. However, there is no compelling evidence that \zgal114 is part of a bound group with \zgal125. In support of this scenario is the similar recessional velocity of the two objects (their velocity difference along the line of sight is $\simeq 70$ \kms). Also, the presence of stripped gas (HII regions) kinematically associated to \zgal114 in the N direction is difficult to explain with a unidirectional ram pressure force acting on the galaxy and might be interpreted as a sign of a gravitational interaction with \zgal125 or its precursors. A similar conclusion was reached by \citet{Merluzzi16}, who studied asymmetries in the gaseous tail of SOS90630, a galaxy in the Shapley Supercluster.  However, the stellar velocity field of \zgal114 appears unperturbed which seems to rule out strong tidal forces produced by \zgal125 on this object. 

\subsection{Is \zgal120 a curiously superimposed object?}

\citet{Cortese06} concluded that it is difficult to believe that the association between \zgal120 and the ionized gas is a mere coincidence. However these authors stated that with the data in hand it was difficult to come to a firm conclusion. In this work we added several pieces of evidence, mostly in support of a pure by-chance superimposition scenario. 

\begin{figure*}
\centering
\includegraphics[width=0.90\textwidth]{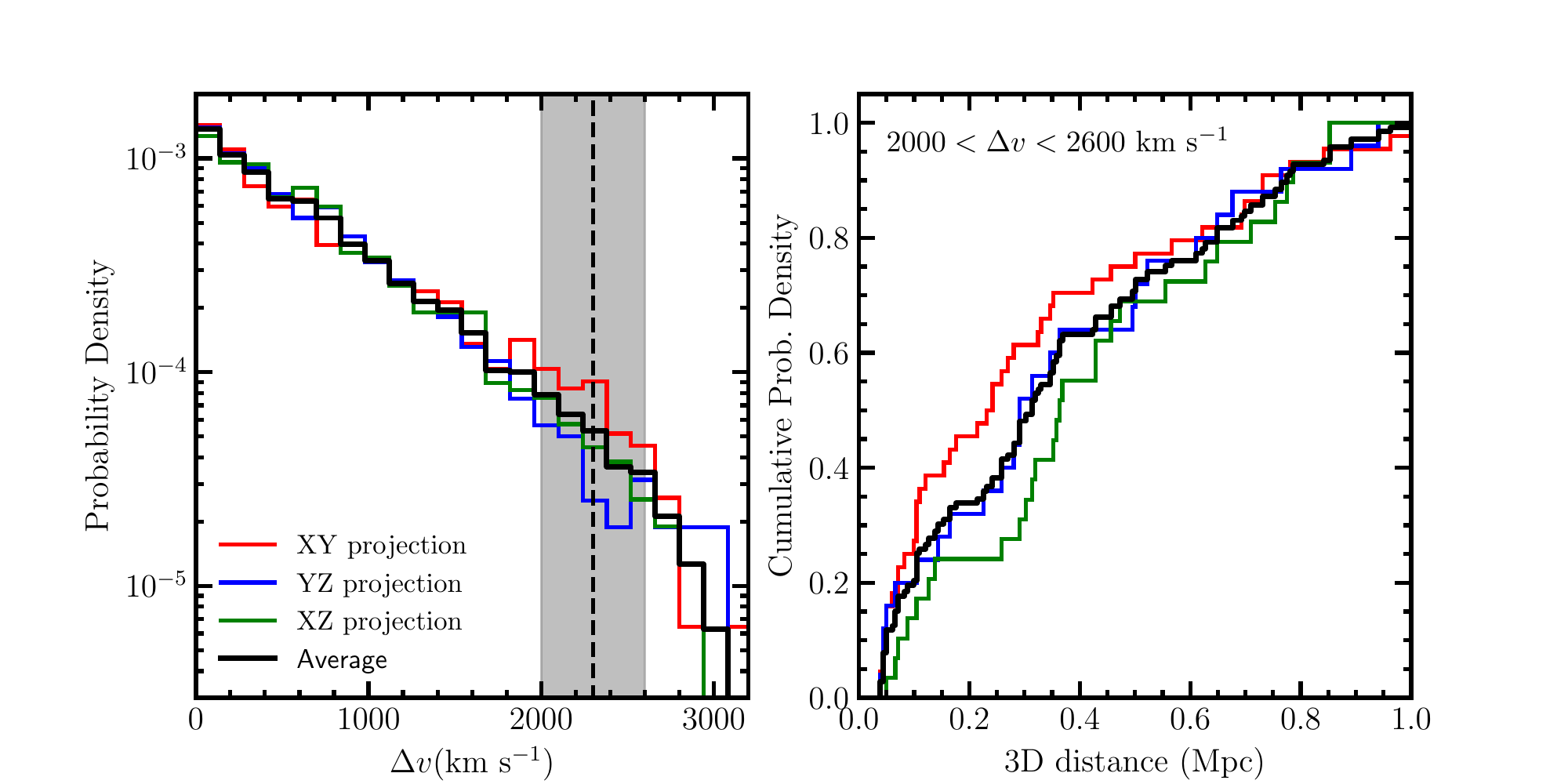}
\caption{Left Panel: probability density distribution of the line of sight velocity offset for pairs of galaxies with stellar masses similar to \zgal120 and \zgal125, separated by $45 \pm 10$ kpc on the sky extracted from a semi analytic model of galaxy formation. The colored lines refer to different projections of the simulation box. The solid black line is the average over the three projections. The vertical dashed line is the observed velocity separation between \zgal120 and \zgal125. Right Panel:
cumulative histogram of the real 3D distance between pairs selected as in the left panel but with the additional constraint that the line of sight velocity offset is in the range $\Delta v = 2000-2600$ \kms. The lines are color coded as in the left panel.}
\label{BIG_SAM}
\end{figure*}

As for \zgal114, we stress that \zgal120 is certainly a massive satellite of the A1367 cluster. The star forming disk truncation alone supports a ram pressure stripping event between the galaxy ISM and the hot intracluster medium. Integral field studies of ram pressure stripped objects had a significant momentum in the last few years. However, most of the studies \citep[see e.g.][]{Merluzzi13, Fumagalli14, Poggianti17} have focused on RPS events in the plane of the sky, thanks to the appearance of long tails of ionized gas, while the observations of \zgal120 are providing us with the best view of RPS along the line of sight. For the first time we have a clear view of the presence of a narrow ring of shocked gas at the radius where stripping is currently ongoing. This shocked gas can then feed the long tails of ionized gas usually detected when the stripping is viewed on the plane of the sky, and it can explain why shocks are visible throughout the entire extent of tails as witnessed in \citet{Fossati16}, \citet{Consolandi17}, and \citet{Poggianti17}. Moreover we clearly see that ram pressure is unable to directly strip the densest molecular clouds or dense gas in spiral arms, and therefore we do see compact sites of star formation outside the stripping radius, as also found in other objects \citep[][]{Abramson14, Fossati18}.

On the other hand, we do not find indications that \zgal120 is associated to the BIG group or to the ionized gas trails. \citet{Cortese06} found unperturbed gas kinematics in the star forming region of the disk, which in this work we confirm with full 2D maps. Even more interesting is the fact that the depth of the MUSE observations allows us to obtain a 2D map of the stellar kinematics extending 15 kpc from the center of \zgal120. While gravitational perturbations are thought to have significant impact at large galactocentric radii, where the stars are less bound, we found no perturbation on the stellar disk up to the radius accessible with our observations. In addition, the gas in the filament (F1) surrounding \zgal120 shows cold dynamics and no indication of shock heating. The low metal to hydrogen line ratios are suggestive of gas at sub-solar metallicity which could hardly originate from a massive galaxy like \zgal120. 
Lastly, an additional piece of evidence against a recent interaction between \zgal120 and the BIG group is given by our reconstruction of the star formation history of the quenched regions of the galaxy. We find that the quenching event at galactocentric radius $5-6$ kpc started about 550 Myr ago and proceeded gradually. This is consistent with a massive galaxy subject to moderate ram pressure as opposed to the nearly instantaneous quenching of NGC4569, a galaxy of similar stellar mass falling on a high velocity radial orbit into the Virgo cluster \citep{Boselli06a, Boselli16}. If the BIG group is falling into A1367 from the N-W direction, {we assume a tangential velocity in the plane of the sky of $800$ \kms \citep{Yagi17}, }and we assume that \zgal120 is moving along the line of sight (from its face-on ram pressure stripping signatures), then we conclude that at the time the stripping started on the outer regions of \zgal120, BIG was  more than 400 kpc away in the North-West direction compared to the current on-sky position of \zgal120. Given the long timescales and high peculiar velocities involved, it is only by chance that today we can observe the proximity on the plane of the sky between \zgal120 and BIG.

Based on this evidence we must conclude that \zgal120 is simply superimposed on the infalling group in the plane of the sky. We resort to numerical simulations and Semi-Analytic Models (SAM) of galaxy formation to investigate how rare is this occurrence in the dense environment of a cluster of galaxies. We take the $z=0$ snapshot of the \citet{Henriques15} SAM, which is based on the Millennium simulation \citep{Springel01}, and we post-process the catalogue as described in \citet{Fossati15} to turn one axis of the simulation box into a redshift, while the other two axes are used as the spatial position on the plane of the sky. We select all the haloes which have a similar mass, within 0.25 dex, as A1367: $\log(M_{\rm halo}/{\rm M_\odot}) = 14.51$ \citep{Hudson10} where $M_{500}$ has been converted to $M_{200}$ using an NFW \citep{Navarro97} profile with concentration equal to 5. We find 1335 such haloes in the simulation box, we then select all the pairs with stellar mass similar to \zgal120 and \zgal125 within 0.25 dex and separated by a projected distance of $45 \pm 10 $ kpc. The left panel of Figure \ref{BIG_SAM} shows the distribution of the line of sight velocity offset for the selected pairs. Lines of different colors correspond to different spatial projections of the simulation box (the redshift axis is always perpendicular to the selected spatial axes). The black solid line is the average over the three projections. It is clear that the observed velocity separation of 2300 \kms\ (vertical dashed line) is a rare event even in the massive potential of a cluster. The fraction of objects in the range $\Delta v = 2000-2600$ \kms\ is $2-4\%$ of the number of selected objects in the three projections. We plot in the right panels of Figure \ref{BIG_SAM} the cumulative distribution of the 3D distance of the selected pairs in this range of $\Delta v$. While the projected distance is set to $45 \pm 10 $ kpc by the selection we find that the distribution of the real distance extends up to 1 Mpc with a median value at 320 kpc. The probability of the selected pairs being physically associated ($d<100$ kpc) is less than 20\%. We therefore conclude that in the simulation framework it is very rare to find a pair like \zgal120 and \zgal125, but these objects do exists and are more likely to be seen in projection rather than being physically associated. On the other hand, if we add the presence of \zgal114 in the selection criteria, and we ask that it is located at $50 \pm 10 $ kpc from \zgal125 and with a velocity offset within 300 \kms, we then find only 2 triplets in the entire simulation box (0.125 Gpc$^3$) with these properties. This does not affect our conclusion, drawn from the two body problem, that based on a selection made with observed parameters, it is unlikely that \zgal120 and \zgal125 are physically associated. 

This argument, combined with the observational evidences presented above places strong constraints on \zgal120 being an object superimposed to the spectacular picture of perturbed galaxies and ionized gas trails of the BIG group.

\subsection{What is the origin of the ionized gas filaments?}

With MUSE we have been able to observe the kinematics and ionization conditions of the ionized gas trails that fill the area of the BIG group. We have also observed the inner part of the $\simeq 330$ kpc tail \citep{Yagi17} extending to the North-West. The filament kinematics is highly complex (see Figure \ref{vel_map}): F1 and F3 appear to have a similar recessional velocity, although F3 shows two distinct filaments in a cross shape with a velocity difference of 200 \kms. The filament F2 instead shows a higher velocity (by 400 \kms\ compared to F1 and F3) with a gradient East-West which can be explained by the acceleration the gas was subject to during the merger event in \zgal125. This kinematics can be qualitatively explained by the complex trajectories that the precursors of \zgal125 might have followed during the merger phase, combined with a time variable tidal stripping of the gas as the galaxies approached the final merger event. 

The filaments appear to be different also in their ionization conditions, F2 shows the highest H$\alpha$ surface brightness and stronger ionization than F1 and F3 {(see Figures \ref{Halpha_map} and \ref{nii_ha_map})}. Indeed F2 is also hosting the largest number of extragalactic HII regions compared to the other two filaments. We recall that in low density diffuse gas ($n_e < 1~{\rm cm^{-3}}$), the recombination time of ionized gas is less than 1 Myr \citep{Fossati16}, while the length of the filaments and the time to reach a post merger morphology in \zgal125 requires there features to live for several tens of Myr.

\citet{Poggianti18} claim that the ram pressure stripped tails of local star forming galaxies can be kept ionized by leakage of ionizing photons from the HII clumps (where they estimate an average escape fraction of 18\%), or lower luminosity HII regions that cannot be individually identified. While this is certainly true, it would require a continuously sustained star formation activity throughout the filaments. This is not implausible, given that \citet{Jachym14} found that the cold molecular gas phase is the most abundant phase (in mass) in the ram pressure stripped tail of ESO137-001. {Moreover \citet{Yamagami11} found that the diffuse molecular gas could be abundant in stripped tails, leading to several episodes of cloud condensation and star formation.} \citet{Cortese06} found atomic neutral gas (HI) in two Arecibo pointings on the BIG group and on the North-West tail. The HI spectra show velocity components which are consistent with the kinematics of the ionized filaments F1 and F3, so we can conclude that the filaments are present in multiple gas phases. 

The elevated [OI]/H$\alpha$ line ratios in all the diffuse gas filaments are also suggestive of other ionization phenomena (beyond stellar photoionization) occurring in the filaments. Among these there are shocks, and thermal mixing with the hot ICM. In summary, the combined effect of these phenomena with photoionization from young stars can explain the long survival time of the filaments. 

\section{Conclusions} \label{sec_conclusions}
In this work we presented a new large mosaic of MUSE observations of the Blue Infalling Group (BIG). Due to their proximity, groups like BIG are ideal targets 
to study the pre-processing scenario \citep{Dressler04}, i.e. the transformation of galaxies from star forming to quiescent in the group environment, prior to their accretion into massive clusters. The depth and full spectral coverage of our observations allowed us to extract resolved stellar velocity and ionized gas maps to study the kinematics and ionization conditions of the gas in the group environment. Our new observations, in combination with existing data from \citet{Cortese06} and \citet{Yagi17}, paint a clearer picture of the processes acting in BIG and in the A1367 cluster environment. 

The group is currently composed of \zgal125, a post-merger galaxy exhibiting a perturbed stellar and gaseous velocity field, and of \zgal114, a lower mass ram pressure stripped galaxy, plus several tidal dwarfs. A few Gyr ago the group was likely outside the virial radius of A1367, on an infall orbit towards the massive cluster. The galaxies that have merged into \zgal125 were orbiting in the group potential, where gravitational and tidal forces have partially stripped them of gas. This interaction combined to the overall motion of the group towards the cluster gave rise to the $\simeq 330$ kpc tail of ionized gas and to the complex pattern of filaments in the center of the group. \zgal114 could also have been subject to gravitational forces, which combined with an overall ram pressure stripping force generated the downstream tail and the star forming blobs on the North side. 
The diffuse gas is kept ionized by a combination of photoionization from {\it in-situ} star formation, shocks, and possibly thermal conduction with the hot cluster ICM. 

Based on new data we conclude that the massive galaxy \zgal120 is not part of BIG but is simply superimposed to the group in the dense cluster environment. No sign of gravitational perturbations is indeed detected on the stellar disk and the diffuse gas surrounding \zgal120 is kinematically relaxed and does not show any sign of shock or turbulence. Comparisons with semi analytic models of galaxy formation showed that this by-chance superposition is rare but possible, and that a velocity difference between \zgal120 and BIG in excess of 2000 \kms\ is likely to imply large physical distances along the line of sight. Instead, \zgal120 gave us one of the best views of face-on ram pressure stripping in a cluster. Our observations revealed a narrow ring of shocked gas at the radius where the stripping is removing the star forming gas. We found that dense molecular clouds can survive the stripping and keep forming stars in a region mostly devoided of gas. A reconstruction of the star formation history of the galaxy outside the stripping radius shows that for this object the stripping is rapid but not instantaneous, a behaviour expected for a massive galaxy.

In conclusion our observations have uncovered that gravitational interactions combined with ram pressure stripping are jointly removing gas from \zgal125 and \zgal114, eventually quenching their star formation activity by the combined effect of the group and cluster environments. These observations show a relevant role for pre-processing in a group like halo in the local Universe. It is likely that a similar combination of physical processes contributed to the onset of environmental quenching of the star formation activity in the early Universe during the formation of the first massive haloes. 

\section*{Acknowledgements}

MFumagalli acknowledges support by the Science and Technology Facilities Council [grant number  ST/P000541/1]. This project has received funding from the European Research Council (ERC) under the European Union's Horizon 2020 research and innovation programme (grant agreement No 757535).
D.J.W. acknowledges the support of the Deutsche Forschungsgemeinschaft via Projects WI 3871/1-1, and WI 3871/1-2.
We thank the anonymous referee for his/her comments which improved the quality of the paper.
This work is based on observations collected at the European Organisation for Astronomical Research in the Southern Hemisphere under ESO programme ID 095.B-0023. 
Based in part on data collected at Subaru Telescope, which is operated
by the National Astronomical Observatory of Japan.
This research made use of Astropy, a community-developed core Python package for Astronomy \citep{Astropy-Collaboration13}. For access to the codes and advanced data products used in this work, please contact the authors or visit \url{http://www.michelefumagalli.com/codes.html}. Raw data are publicly available via the ESO Science Archive Facility.




\bibliographystyle{mnras}

  






\bsp 
\label{lastpage}
\end{document}